\RequirePackage{fixltx2e}[2005/12/01] 
\documentclass[useAMS,usenatbib]{mn2e}
\pdfoutput=1
\usepackage{amsmath}
\usepackage[varg]{txfonts}
\usepackage{astrojournals}
\usepackage{graphicx}
\usepackage{microtype}
\usepackage{xcolor}
\usepackage{hyperref}
\hypersetup{colorlinks=True, linkcolor=blue!50!black, citecolor=black,
  urlcolor=blue!50!black}

\newcommand\NEW[1]{#1}
\setkeys{Gin}{width=\linewidth}

\newcommand\kms{\ensuremath{\mathrm{km\ s^{-1}}}}

\newcounter{ionstage}
\DeclareRobustCommand\ion[2]{\setcounter{ionstage}{#2}%
  \ensuremath{\text{#1\,\scalebox{1.0}[0.85]{\Roman{ionstage}}}}}
\newcommand\hii{\texorpdfstring{\ion{H}{2}}{H II}}
  
\title[Radiation-MHD simulations of \hii{} regions and
PDRs]{Radiation-magnetohydrodynamic simulations of \hii{} regions and
  their associated PDRs in turbulent molecular clouds} 
\author[S.~J. Arthur et al.]{S. J. Arthur$^{1}$\thanks{E-mail:
    \url{j.arthur@crya.unam.mx}; \url{w.henney@crya.unam.mx}; \url{garrelt@astro.su.se};
  \url{fabio@ucolick.org}; \url{e.vazquez@crya.unam.mx}}, W. J. Henney$^{1}$, G. Mellema$^{2}$,
  F. De Colle$^{3}$ and E. V\'azquez-Semadeni$^{1}$\\
$^1$Centro de Radioastronom\'{\i}a y Astrof\'{\i}sica, Universidad Nacional Aut\'onoma de M\'exico,
  Campus Morelia, Apdo. Postal 3-72, 58090 Morelia, Michoac\'an,
  M\'exico\\
$^2$Department of Astronomy \& Oskar Klein Centre, AlbaNova, Stockholm
University, SE 10691 Stockholm, Sweden\\
$^3$Department of Astronomy and Astrophysics, University of California Santa Cruz, Santa Cruz, CA 95064, USA
}

\begin{document}


\maketitle

\begin{abstract}
  We present the results of radiation-magnetohydrodynamic simulations
  of the formation and expansion of \ion{H}{2} regions and their
  surrounding photodissociation regions (PDRs) in turbulent,
  magnetised, molecular clouds on scales of up to 4 parsecs. We
  include the effects of ionising and non-ionising ultraviolet
  radiation and x~rays from population synthesis models of young star
  clusters. For all our simulations we find that the \ion{H}{2} region
  expansion reduces the disordered component of the magnetic field,
  imposing a large-scale order on the field around its border, with
  the field in the neutral gas tending to lie along the ionisation
  front, while the field in the ionised gas tends to be perpendicular
  to the front. The highest pressure compressed neutral and molecular
  gas is driven towards approximate equipartition between thermal,
  magnetic, and turbulent energy densities, whereas lower pressure
  neutral/molecular gas bifurcates into, on the one hand, quiescent,
  magnetically dominated regions, and, on the other hand, turbulent,
  demagnetised regions. The ionised gas shows approximate
  equipartition between thermal and turbulent energy densities, but
  with magnetic energy densities that are 1 to 3 orders of magnitude
  lower. A high velocity dispersion (\(\sim 8\)~km~s\(^{-1}\)) is
  maintained in the ionised gas throughout our simulations, despite
  the mean expansion velocity being significantly lower. The magnetic
  field does not significantly brake the large-scale \ion{H}{2} region
  expansion on the length and timescales accessible to our
  simulations, but it does tend to suppress the smallest-scale
  fragmentation and radiation-driven implosion of neutral/molecular
  gas that forms globules and pillars at the edge of the \ion{H}{2}
  region. However, the relative luminosity of ionising and
  non-ionising radiation has a much larger influence than the presence
  or absence of the magnetic field. When the star cluster radiation
  field is relatively soft (as in the case of a lower mass cluster,
  containing an earliest spectral type of B0.5), then fragmentation is
  less vigorous and a thick, relatively smooth PDR forms. \textbf{Accompanying
  movies are available at \url{http://youtube.com/user/divBequals0}}
\end{abstract}


\section{Introduction}
\label{sec:intro}
\ion{H}{2} regions, that is, regions of photoionised gas, are among
the most arresting astronomical objects at optical wavelengths. The
basic theory behind their formation and expansion has been known for
some time
\citep{{1939ApJ....89..526S},{1954BAN....12..187K},{1978ppim.book.....S}}. More
recently, attention has turned to explaining the irregular structures,
filaments, globules and clumps seen within and around these
regions. These could be due to underlying density inhomogeneities in
the molecular cloud into which the \ion{H}{2} region is expanding, or
could be due to instabilities at the ionisation front itself
\citep{{1996ApJ...469..171G},{2006ApJ...647..397M},{2001MNRAS.327..788W},{2010ApJ...723..971G},{2010MNRAS.403..714M}}.  The effect of the
stellar ionising radiation on these structures can lead to radiation
driven implosion, and the subsequent compression could result in the
formation of new stars \citep{{1989ApJ...346..735B},{2007MNRAS.377..383E},{2007A&A...467..657M}}.

Magnetic fields are well known to pervade molecular clouds and are
thought to play an important role in regulating the star-formation
process by providing support to collapsing structures in molecular
clouds \citep{1999osps.conf..305M}, although recently their relative importance in this process has become less clear \citep{2007ARAA...45..565M}. However, theoretical work on the
interplay between ionising radiation and magnetic fields in and around
\ion{H}{2} regions has only been possible recently, with the advent of
radiation-magnetohydrodynamic codes. \citet{2007ApJ...671..518K}
performed the first simulations of the expansion of an \ion{H}{2}
region in a uniform, magnetised medium. At early times, the thermal
pressure of the photoionised gas dominates over the magnetic field,
which is pushed out of the expanding \ion{H}{2} region. At late times
(several Myr), the photoionised region becomes elongated along the field
lines and the magnetic field filled back
in. \citet{2009MNRAS.398..157H} and \citet{2010arXiv1012.1500M} have
studied the effects of ionising radiation on magnetised globules,
finding that the photoevaporation process is significantly altered
only when the magnetic pressure exceeds the thermal pressure by more
than a factor of ten in the initial
globule. \citet{2010arXiv1010.5905P} have recently studied the
combined effects of magnetic fields and photoionisation on the
formation of a massive star on scales \(< 0.1\)~pc, finding that
rotational winding of the field lines produces a magnetised `bubble',
whose magnetic pressure acts to confine the nascent \hii{}
region during its ultracompact phase. 

In this paper, we study the expansion of \ion{H}{2} regions in
non-uniform (turbulent) magnetised molecular clouds. Previously, we
studied \ion{H}{2} region expansion in a turbulent medium without a
magnetic field and our results were strikingly similar to observed
\ion{H}{2} regions \citep{2006ApJ...647..397M}. In the present work,
we study the effect of the expanding photoionised gas on structures in
the surrounding magnetised medium and also the effect that the
turbulent magnetic field has on the \ion{H}{2} region itself. At least
at the relatively early times studied in our simulations, most of the
interesting effects due to the magnetic field will occur within the
neutral, photon dominated region around the \ion{H}{2} region, and so
we use a careful treatment of the heating and cooling processes in
this gas, first discussed in \citet{2009MNRAS.398..157H}, in order to
have a realistic portrayal of the dynamical processes here.

The wealth of new observational data at infrared wavelengths enables
comparisons to be made of our simulations not only with observations
of the photoionised gas, as done previously in
\citet{2006ApJ...647..397M} but also with the PDR and molecular gas
affected by the \ion{H}{2} region. In particular, the Spitzer GLIMPSE
surveys of bubbles in the Galactic plane
\citep{{2006ApJ...649..759C},{2007ApJ...670..428C}} highlight polycyclic aromatic hydrocarbon (PAH)
emission at the outer edges of \ion{H}{2} regions and in PDRs. These
surveys show that the thickness of the PAH-emitting shell varies
between 0.2 and 0.4 of the outer radius and that shell thickness
increases approximately linearly with radius. Moreover, the bubbles
are generally non-spherical. Herschel and APEX observations reveal
thermal emission from cold dust in the molecular gas outside the PDR
\citep[e.g.][]{{2010A&A...518L..99A},{2010A&A...523A...6D}}. These
observations are useful in identifying condensations that could be
collapsing to form new stars.

Studies of magnetic fields in and around molecular clouds are
important for understanding the star-formation process. \NEW{There
is no consensus as to the importance of magnetic fields in the
formation of clouds, cores, and ultimately stars. If magnetic fields
are dynamically important, then star formation is regulated by
ambipolar diffusion
\citep[e.g.,][]{1999osps.conf..305M}. Alternatively, if magnetic
fields are unimportant then other processes such as turbulence and
stellar feedback provide support to molecular clouds and determine the
star-formation efficiency (e.g.,
\citealp{{1999ApJ...515..286B},{2000ApJ...530..277E},{2002ApJ...566..302M},{2006ApJ...653..361K},{2009ApJ...700..358T},{2011arXiv1101.1534T}}).  It is not
known whether magnetic fields are strong enough to dynamically support
molecular clouds and more evidence is needed. It is possible that
reality lies somewhere between these extremes. Recent Zeeman
observations of molecular clouds
\citep{{2009ApJ...692..844C},{2010ApJ...725..466C}} do not agree with
predictions of idealized ambipolar diffusion models (strong magnetic
field) but more observations are needed.}


There are three main
techniques for tracing the magnetic field in diffuse HI and molecular
clouds: polarisation from aligned dust grains, linear polarisation of
spectral lines, and Zeeman splitting of spectral lines
\citep{2005LNP...664..137H}. Polarisation studies yield $B_\perp$, the
strength of the field projected onto the plane of the sky, whereas
Zeeman splitting gives $B_{||}$, the strength of the field parallel to
the line of sight. In particular, the \ion{H}{1} Zeeman effect has
been used to study the atomic gas (assumed to be in the
photodissociated region) in the star-forming regions M17
\citep{2001ApJ...560..821B} and the Ophiuchus dark cloud complex
\citep{1994ApJ...424..208G}, while the OH Zeeman effect is used to
study the line-of-sight magnetic field in the molecular gas. The
results of such studies suggest that there is rough equipartition of
magnetic and turbulent (dynamical) energy densities in the neutral
gas. In the Ophiuchus dark cloud region, the uniform field in the
\ion{H}{1} gas is estimated to be 10.2~$\mu$G
\citep{1994ApJ...424..208G}, whereas the magnitude of the
line-of-sight magnetic field strength in the \ion{H}{1} gas associated
with M17 is found to be $\sim 500 \mu$G (at $60\arcsec$ resolution, or
even twice as high at $26\arcsec$ resolution)
\citep{2001ApJ...560..821B}, suggesting that there is small-scale
structure in the magnetic field. Recent linear polarisation studies at
submillimetre wavelengths of the magnetic field around the
ultracompact \ion{H}{2} region G5.89--0.39
\citep{2009ApJ....695.1399T} provide evidence for compression of the B
field in the surrounding molecular cloud and general disturbance of
the magnetic field caused by the expansion of the \ion{H}{2} region.

For photoionised gas, Faraday rotation measures of line-of-sight
extragalactic radio sources are used to calculate the magnetic field
strengths in foreground \ion{H}{2} regions
\citep{1981ApJ...247L..77H}. The Zeeman effect cannot be used in the
ionised gas because the thermal linewidths are too
large. \citet{1981ApJ...247L..77H} use double extragalactic radio
sources to sample two positions in the \ion{H}{2} regions S117 and
S264 and find identical rotation measures in each case of order 1 or
$2\,\mu$G. In these regions, the thermal energy density dominates the
magnetic energy density. In a third \ion{H}{2} region, S119, for which
only a single rotation measure is available, the line-of-sight
magnetic field strength is around 20~$\mu$G and the ratio of magnetic
to thermal energy density is around 0.4, hence the magnetic field
could have had an effect on the expansion of this \ion{H}{2} region.

The remainder of this paper is organised as follows. 
In \S~\ref{sec:model} we describe our numerical algorithm and present
two test problems: the classical Spitzer law for the expansion of a
non-magnetised \hii{} region in a uniform medium, and the expansion of
an \ion{H}{2} region in a uniform magnetised medium
\citep{{2007ApJ...671..518K},{2010arXiv1012.1500M}}. In the second
case, we draw attention to some interesting properties of the solution
that have not been described previously.

In \S~\ref{sec:turbhii} we describe the expansion of \ion{H}{2}
regions in both magnetised and non-magnetised turbulent media, taking
as initial conditions for the ambient medium the results of an MHD
turbulence calculation \citep{2005ApJ...618..344V}, and considering
both a strong (approximately O9 spectral type) ionising source and a
weak (B0.5) one. In \S~\ref{sec:discussion} we make qualitative
comparisons between simulated optical and long-wavelength images and
observations. We also produce synthetic maps of the projected
line-of-sight and plane-of-sky components of the magnetic field, which
can be qualitatively compared with observations of the magnetic field
in star-forming regions. We discuss the globules and filaments at the
periphery of our simulated \ion{H}{2} regions in the context of recent
numerical work on photoionised magnetised globules
\citep{{2009MNRAS.398..157H}, {2010arXiv1012.1500M}}. Finally, in
\S~\ref{sec:conclusion} we outline our conclusions.

\section{Numerical Model}
\label{sec:model}
\subsection{Numerical Method}
\label{sec:nummethod}
The numerical method used in this paper is the same as that described
by \citet{2009MNRAS.398..157H}. The Phab-C$^2$ code combines the
ideal magnetohydrodynamic code described by \citet{2006A&A...449.1061D} with
the C$^2$-Ray method (Conservative Causal Ray) for radiative transfer
developed by \citet{2006NewA...11..374M}, and adds a new treatment of
the heating and cooling in the neutral gas
\citep{2009MNRAS.398..157H}. The MHD code is a second-order upwind
scheme that integrates the ideal MHD equations using a Godunov method with a
Riemann solver, similar to that described by
\citet{1998MNRAS.297..265F} but with the constrained transport (CT)
method (e.g., \citealp{2000JCoPh.161..605T}) used to maintain the divergence
of the magnetic field close to zero. An artificial (Lapidus) viscosity
is also included \citep{1984JCoPh..54..174C} in order to broaden the
discontinuities and reduce oscillations. The C$^2$-Ray method for
radiative transfer is explicitly photon conserving
\citep{2006NewA...11..374M} and the algorithm allows for timesteps
much longer than the characteristic ionisation timescales or the
timescale for the ionisation front to cross a numerical grid cell, by
means of an analytical relaxation solution for the ionisation rate
equations. This results in a highly efficient radiative transfer
method.

The MHD and radiative transfer codes are coupled via operator
splitting, the inclusion of equations for the advection of ionised and
neutral density in the MHD code, and the energy equation, which
includes both radiative cooling and heating due to the absorption of
stellar radiation by gas or dust: ionising extreme ultraviolet
radiation in the photoionised region, non-ionising far ultraviolet
radiation in the neutral gas and x-rays in the dense molecular gas. A
detailed discussion of the heating and cooling contributions is given
by \citet{2009MNRAS.398..157H}. In particular, the treatment of
heating in the neutral PDR region, taking into account FUV/optical
dust heating and x-ray heating, is a substantial improvement over
standard treatments in the literature. The heating term for the
neutral gas is calibrated using Cloudy \citep{1998PASP..110..761F} and
is tailored to the ionisation source. Moreover, x-rays and FUV
radiation will be enhanced due to the presence of associated low-mass
stars in the star-formation environment, and we use the results of
\citet{2008ApJ...675.1361F} to estimate the distributions of FUV and
EUV fluxes appropriate to stellar clusters of membership sizes
relevant to the two cases we study: an O9 and a B0.5 star. The energy
equation is solved iteratively using substepping and the timestep for
the whole calculation is determined solely by the magnetohydrodynamic
timestep.

The individual components of the Phab-C$^2$ code have been tested
against standard problems
\citep{{2005PhD...........F},{2004Ap&SS.293..173D},{2006MNRAS.371.1057I}}
and the whole code was applied to the problem of the photoionisation
of magnetised globules \citet{2009MNRAS.398..157H}, work which has
recently been independently corroborated by
\citet{2010arXiv1012.1500M}. In this paper, we have implemented an
entropy fix \citep{1999JCoPh.148..133B} to correctly deal with the
thermal and dynamical pressures in regions of low density where the
magnetic pressure dominates.  We have also improved the treatment of
the boundary conditions because they can become important,
particularly towards the end of the simulations. For our O star
simulations we use outflow boundary conditions, since thermal pressure
is expected to dominate throughout the evolution and the \ion{H}{2}
region will rapidly grow to the size of the computational box. For the
B star simulations, the expansion timescale of the \ion{H}{2} region
is much longer and is comparable to the crossing time of the
computational box corresponding to the initial turbulent velocity
dispersion of the molecular gas. We therefore use periodic boundary
conditions in this case, which ensures that the total mass in the
computational box remains constant.

Parallelisation using Open-MP was implemented where possible, enabling
the current simulations to be run in about two weeks on an Intel
8-core server with 32~GB of RAM. Preliminary work was performed on the Kan
Balam supercomputer of the Universidad Nacional Aut\'onoma de M\'exico.

\subsection{Test Problems}
\label{sec:testprob}
\subsubsection{Hydrodynamic \hii{} Region Expansion in a Uniform Medium}
\label{sec:nomhd}
\begin{figure}
\centering
  \includegraphics[width=\linewidth]{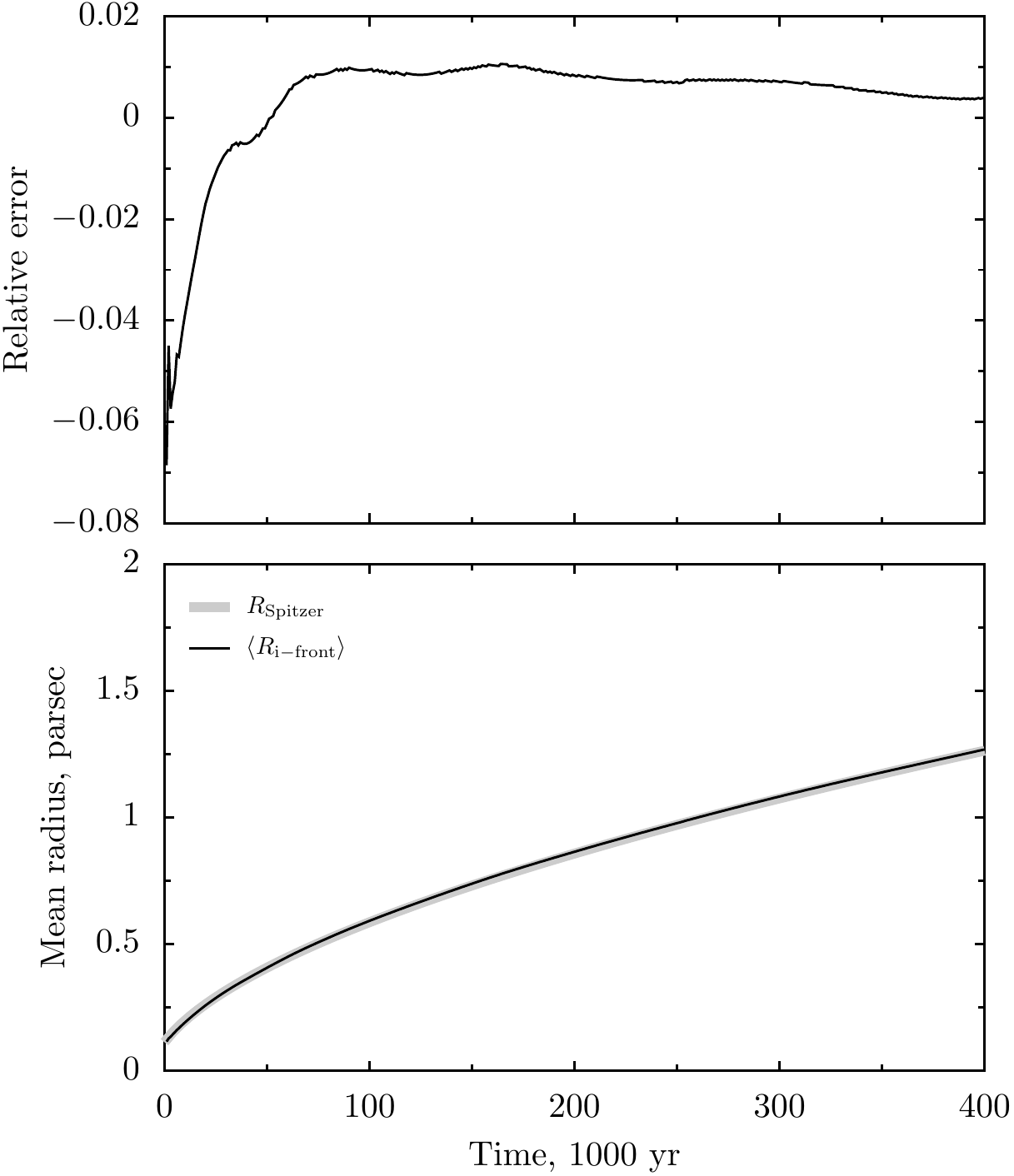}
\caption{Radius against time for the hydrodynamic expansion of a
  photoionised region in a uniform medium. Bottom panel: numerical
  simulation (black line) and analytical solution (gray line). Top
  panel: relative error. }
\label{fig:nomhd_rad}
\end{figure}Org/
\begin{figure}
\centering
  \includegraphics[width=\linewidth]{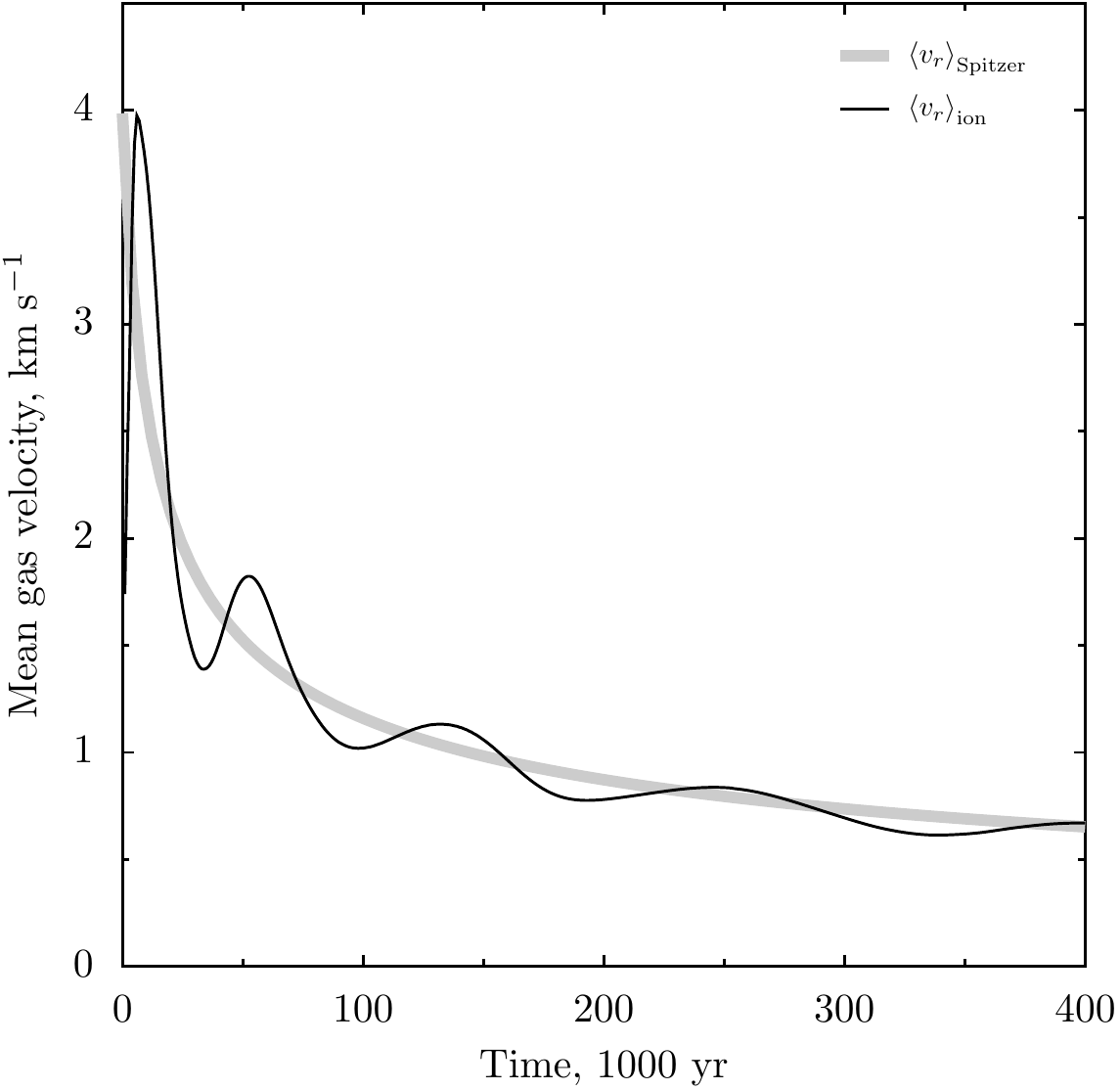}
\caption{Mean gas velocity against time for the hydrodynamic expansion
of a photoionised region in a uniform medium. Black line --- numerical
simulation; thick gray line --- analytic solution.}
\label{fig:nomhd_vel}
\end{figure}

The radiation part of the code has been extensively tested and
documented \citep{{2006NewA...11..374M},{2006MNRAS.371.1057I}}. In
order to test the radiation (magneto-)hydrodynamics combined code we
first consider the purely hydrodynamic evolution of a photoionised
region around an ionising source in a uniform medium. The analytical
solution to this problem is the familiar \citet{1978ppim.book.....S}
expansion law
\begin{equation}
  R_\mathrm{Spitzer} = R_0\left(1 + \frac{7}{4}\frac{c_\mathrm{i} t}{R_0}\right)^{4/7} \ ,
\label{eq:spitz}
\end{equation}
where $R_\mathrm{Spitzer}$ is the ionisation front radius, $R_0 = (3
Q_\mathrm{H}/4\pi n_0^2 \alpha_\mathrm{B})^{1/3}$ is the initial
Str\"omgren radius, $c_\mathrm{i}$ is the sound speed in the ionised
gas and $t$ is time, $Q_\mathrm{H}$ is the ionising photon rate, $n_0$
is the ambient density and $\alpha_\mathrm{B}$ is the case B
recombination rate, $\alpha_\mathrm{B} = 2.6\times 10^{-13} T^{-0.7}$
\citep{2006agna.book.....O}, where $T$ is the temperature in the
ionised gas. For an ambient density $n_0 = 1000$~cm$^{-3}$, stellar
ionising photon rate $Q_\mathrm{H} = 5 \times 10^{46}$~s$^{-1}$ and
stellar temperature $T_\mathrm{eff} = 37500$~K the results are shown
in Fig.~\ref{fig:nomhd_rad}, together with the corresponding analytic
solution. These parameters were chosen because they represent the mean
values of the turbulent medium B star case described below in
\S~\ref{sec:turbhii}. The lower panel shows radius against time, while
the upper panel shows the relative error, defined as
\begin{equation}
  \mbox{Error} = \frac{R_\mathrm{ion} - R_\mathrm{Spitzer}}{R_\mathrm{Spitzer}} \ .
\end{equation}
From this figure, we see that the largest discrepancies, of about 6\%,
occur at the beginning of the expansion, when a shock wave begins to
propagate outwards ahead of the ionisation front and a corresponding
rarefaction wave travels back towards the ionising source, which is
not accounted for in the analytical solution. At later times, the
error is never more than 1\%. Our code, therefore, compares extremely
favorably with other radiation-hydrodynamics codes on this problem
\citep[see, e.g.][]{2006ApJS..165..283A, 2007ApJ...671..518K,
  2009MNRAS.400.1283I}. There is no evidence for overcooling in the
ionisation front in our simulations, as was found to be a problem by \citet{2007ApJ...671..518K}. In the analytic Spitzer solution we
assume an \hii{} region temperature \(T = 8900\)~K, which gives the
best fit to our numerical result. This is about \(400\)~K hotter than
the volume-averaged ionised gas temperature in our simulation, but is
roughly equal to the ionised gas temperature just inside the
ionisation front. 

In Fig.~\ref{fig:nomhd_vel} we plot the gas radial velocity from the
numerical simulation and the corresponding analytical solution. The
numerical solution closely follows the analytical solution, though
with oscillations due to acoustic phenomena, such as the initial
rarefaction wave, which propagates back into the \ion{H}{2} region
when the ionization front begins to expand outwards.

\subsubsection{MHD \hii{} Region Expansion}
\label{sec:krum}
\begin{figure*}
\centering
  \includegraphics[width=\linewidth]{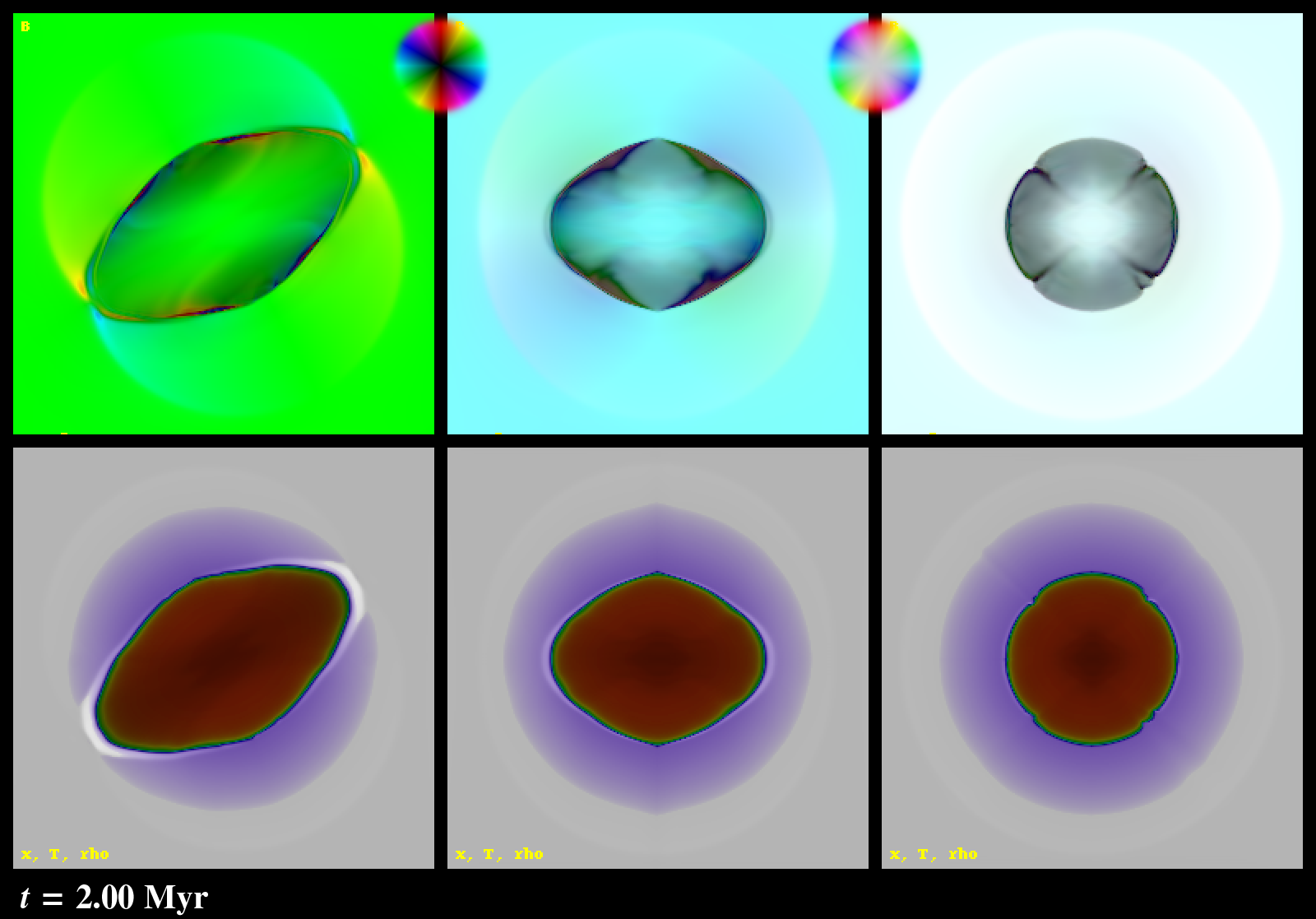} 
  \caption{Expansion of an \ion{H}{2} region in a magnetised uniform
    medium after 2~Myr in a $(20\,\mathrm{pc})^3$ computational box. The top row
    shows the magnetic field in cuts in the central $x$-$y$ (left),
    $x$-$z$ (centre) and $y$-$z$ (right) planes. The colour indicates
    the direction of the magnetic field, where the coloured discs show
    the field angle in the plane for a strong field (left, brighter
    disc) and a weaker field (right, dimmer disc). Grey indicates
    magnetic field perpendicular to the plane. The bottom row shows
    ionisation fraction $x$, temperature \(T\) and density $\rho$,
    coded as `hue', `saturation', and `value', respectively. Red
    indicates hot ($\sim 10^4$~K) ionised gas, blue/green is partially
    ionised gas, purple is warm ($\sim 300$~K) neutral (PDR) gas and
    grey indicates cold neutral (molecular) gas. The density is
    represented by the intensity, with the densest gas being white and
    the most diffuse is black.}
\label{fig:krum1}
\end{figure*}
\begin{figure*}
\centering
  \includegraphics[width=\linewidth]{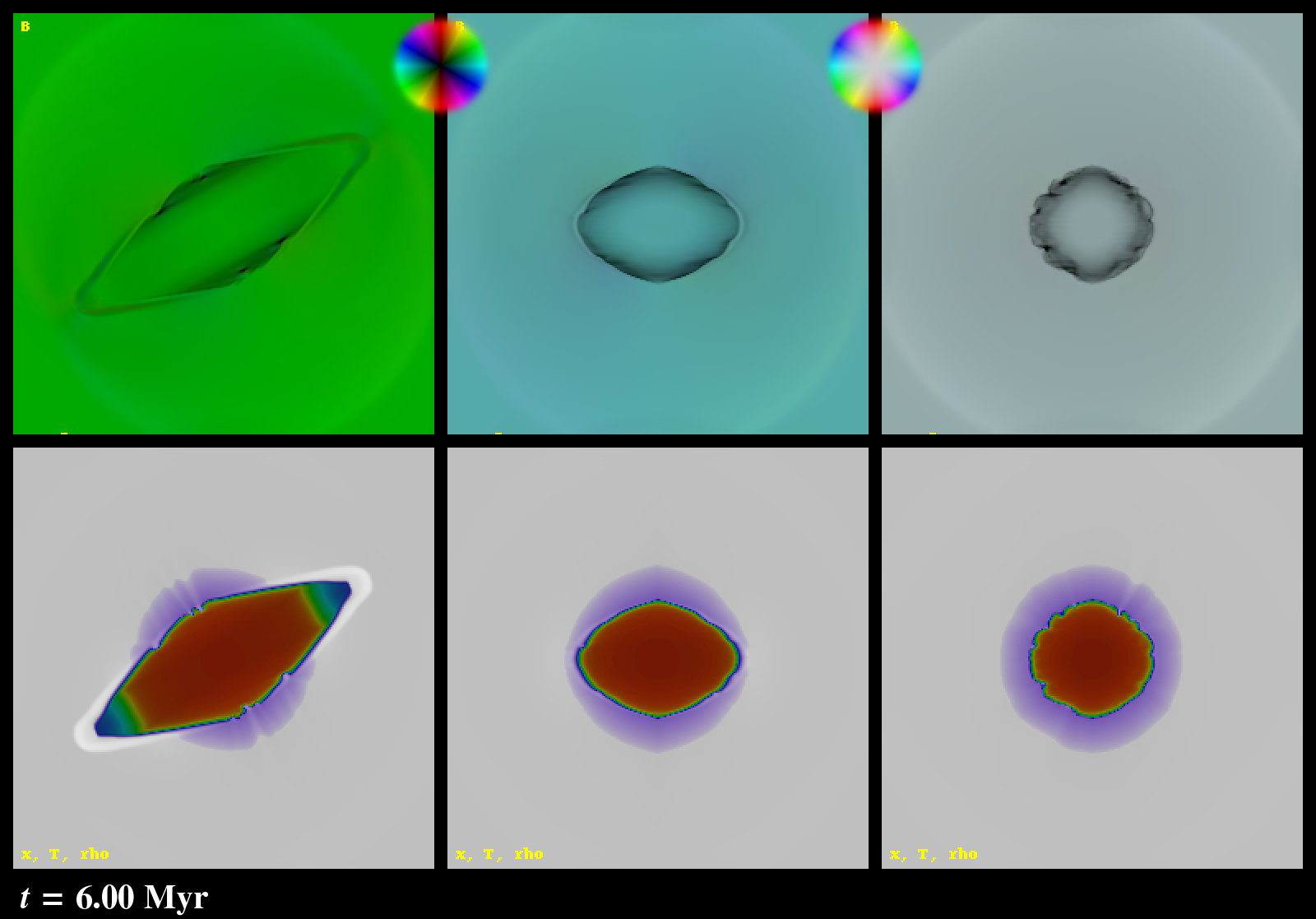}
\caption{Same as Fig.~\protect\ref{fig:krum1} but after 6~Myr in a
  $(40\,\mathrm{pc})^3$ computational box.}
\label{fig:krum2}
\end{figure*}
There is no corresponding analytical solution for the expansion of an
\ion{H}{2} region into a magnetised medium. Therefore, in order to
test the MHD part of the numerical code, we carry out a simulation of
the expansion of an \ion{H}{2} region in a uniform magnetised medium
using the same parameters as those of
\citet{2007ApJ...671..518K}. This problem was also used to test the
radiation MHD code of \citet{2010arXiv1012.1500M}. The simulation is
performed in a $256^3$ computational box with constant ambient density
and temperature of $n_0 = 10^2$~cm$^{-3}$ and $T_0 = 11$~K,
respectively, a constant ambient magnetic field directed at $30^{\circ}$
to the $x$-axis in the $x$-$y$ plane\footnote{The original simulation
performed by \citet{2007ApJ...671..518K} has the magnetic field
parallel to the $x$-axis.} of strength $B_0 = 14.2\, \mu$G and an
ionising source producing $Q_\mathrm{H} = 4 \times 10^{46}$~ionising
photons~s$^{-1}$. The magnetic field is placed at an angle to the
grid lines in order to be able to distinguish between physical and
computational effects. 
\NEW{We run the simulation in a $(20\,\mathrm{pc})^3$ box and in a
$(40\,\mathrm{pc})^3$ box, the former to study the early evolution $(t
< 2$~Myr) and the latter to highlight the late-time evolution.}
We note that the ionising source is very weak
\citep[later than spectral type B0.5,][]{1996ApJ...460..914V} and that
the ambient density is lower than that typically found in star-forming
regions.

\citet{2007ApJ...671..518K} found a critical radius and time for when
magnetic pressure and tension start to become important for the
expansion of the \ion{H}{2} region in a uniform magnetised medium,
which corresponds to when the magnetic pressure in the neutral gas
becomes of the same order as the thermal pressure in the ionised gas,
i.e., $\rho_\mathrm{n}v_\mathrm{A}^2 \sim \rho_\mathrm{i} c_\mathrm{i}^2$. Here, $\rho_\mathrm{n}$, $\rho_\mathrm{i}$ are
the densities in the neutral and ionised gas, $v_\mathrm{A}$ is the Alfv\'en
speed, and $c_\mathrm{i}$ is the sound speed in the ionised gas. Thus, magnetic
effects become significant when the \ion{H}{2} region has expanded to
a radius of roughly
\begin{equation}
R_\mathrm{m} \equiv \left( \frac{c_\mathrm{i}}{v_\mathrm{A}}\right)^{4/3} R_0 \ ,
\label{eq:rm}
\end{equation}
where $R_0$ is the initial Str\"omgren radius in a non-magnetised
medium of the same uniform density. The corresponding time
is
\begin{equation}
  t_\mathrm{m} \equiv \frac{4}{7} \left( \frac{c_\mathrm{i}}{v_\mathrm{A}} \right)^{7/3} t_0 \ ,
\label{eq:tm}
\end{equation}
where $t_0$ is the sound crossing time of the initial Str\"omgren
sphere, and both of these equations implicitly assume that the
classical Spitzer expansion law \citep{1978ppim.book.....S} holds
until the magnetic field becomes important.

For the parameters in this test problem, the initial Str\"omgren
radius is $R_0 = 0.48$~pc, and the Alfv\'en speed in the ambient
magnetised medium is $2.6$~km~s$^{-1}$. Our simulations give a
temperature in the photoionised gas of $T_\mathrm{i} \sim 9000$~K and hence a
sound speed of $c_\mathrm{i} = 9.8$~km~s$^{-1}$ (compare the lower assumed
ionised sound speed $c_\mathrm{i} \sim 8.7$~km~s$^{-1}$ in the
\citealt{2007ApJ...671..518K} models). Thus, the critical radius and
time are $R_\mathrm{m} = 2.8$~pc and $t_\mathrm{m} = 0.61$~Myr, respectively, for our
models.

In Figs.~\ref{fig:krum1} and \ref{fig:krum2} we show snapshots of the
magnetic field and the photoionised gas distribution in the central
$x$-$y$, $x$-$z$ and $y$-$z$ planes of the computational box after 2
and 6~Myr of evolution.  The coloured discs shown between the panels
of the top row are keys to the interpretation of the magnetic field
images. The hue indicates the projected field orientation in the plane
of each cut (red for vertical, cyan for horizontal, etc.), while the
brightness indicates the magnetic field strength, as is shown in the
left-hand disc. The colour saturation diminishes as the out-of-plane
component of the field increases, as shown in the right-hand disk.
This non-standard representation allows the magnetic field to be
sampled at every pixel.
The lower
panels of the figures show the ionisation fraction, temperature and
density of the gas in the same central planes. Ionisation fraction is
represented by colour: red indicates fully ionised gas while
blue/green is partially ionised. The gas temperature varies beween
$\sim 10^4$~K (red) in the ionised gas to a few hundred Kelvin in the
neutral (PDR) gas. The gas density is represented by the intensity in
these figures, with the densest, cold molecular gas appearing white.

At early times, $t < t_\mathrm{m}$, the \ion{H}{2} region is
essentially spherical and expands at approximately the same speed
(half the sound speed) in all directions because the ionised gas
thermal pressure is much higher than the ambient magnetic pressure.
Expansion across the magnetic field lines compresses the B-field. The
expanding \ion{H}{2} region and PDR are preceded by a fast-mode shock,
which travels furthest perpendicular to the magnetic field
direction. The fast-mode shock bends and compresses the field lines
and so the PDR and \ion{H}{2} region propagate in a modified magnetic
field. The slow-mode shock is strongest along the field lines and most
evident at the end caps, where it produces a dense shell. It
compresses the gas and demagnetises it (reduces the field). The
slow-mode shock is almost isothermal and the dense shell behind it is
pushed by the pressure in the photoionised region. At early
times the magnetic field is reduced in the photoionised gas by the
expansion of the \ion{H}{2} region. At later times, $t > t_\mathrm{m}$, the
magnetic tension pulls the field back into the
\ion{H}{2} region at the equator. 

In Fig.~\ref{fig:krum1} we show the situation after 2~Myr, when the
magnetic field has started to pull back into the \ion{H}{2}
region. The \ion{H}{2} region is elongated along the magnetic field
direction but the photon dominated region (warm neutral gas) is
approximately spherical. The magnetic field retains its original
direction except around the edge of the \ion{H}{2} region. This is
particularly evident at the poles of the \ion{H}{2} region. At the time shown
in the figure, the magnetic field appears concentrated towards the
centre, with regions of lower magnetic field around the equator or
waist of the structure. Neutral gas is pulled into the \ion{H}{2}
region along with the magnetic field, becomes ionised and flows away
from the ionisation front into the \ion{H}{2} region. It then flows
along the field lines and is channelled out towards the poles, where
the ionisation front transforms into a recombination front. We note
that at this time in the simulation, we do see instabilities form in
the direction perpendicular to the magnetic field, as reported by
\citet{2007ApJ...671..518K} and \citet{2010arXiv1012.1500M}. These
instabilities form at the equator where the ionisation front is
parallel to the magnetic field direction because the slow-mode shock
 which precedes the ionisation front, is not permitted to cross the
 field lines. As a result, the ionisation front corrugates.

Our Fig.~\ref{fig:krum2} shows the simulation after 6~Myr in a
$(40\,\mathrm{pc})^{3}$ computational box. The \ion{H}{2} region has stopped
expanding perpendicular to the magnetic field but the shocked neutral
shell is still expanding along the field lines, leading to a very
elongated structure. However, in the direction along the field lines,
the flow has become one dimensional and so the density does not fall
off as the \ion{H}{2} region expands. Thus, there is a limit as to how
far the ionisation/recombination front can travel in this
direction. There is an extended region of partially ionised gas beyond
the recombination front. The warm, neutral gas (PDR) is still roughly
spherical in shape but is pierced by the cigar-shaped \ion{H}{2}
region along the field direction. The dense neutral knots formed as a
result of the instability at the equator shadow parts of the PDR. The
knots formed closest to the perpendicular direction remain there
throughout the simulation, pinned by the magnetic field. The knots
slightly off the equator experience the rocket effect due to the
photoionisation of their outer skins. They move like beads on a wire
along the magnetic field lines around the edge of the \ion{H}{2}
region, a process which can be clearly observed in animations of this
simulation. By the end of the calculation, the magnetic field has
refilled the \ion{H}{2} region and become roughly uniform and returned
completely to its original direction.

\begin{figure}
\centering
  \includegraphics[width=\linewidth]{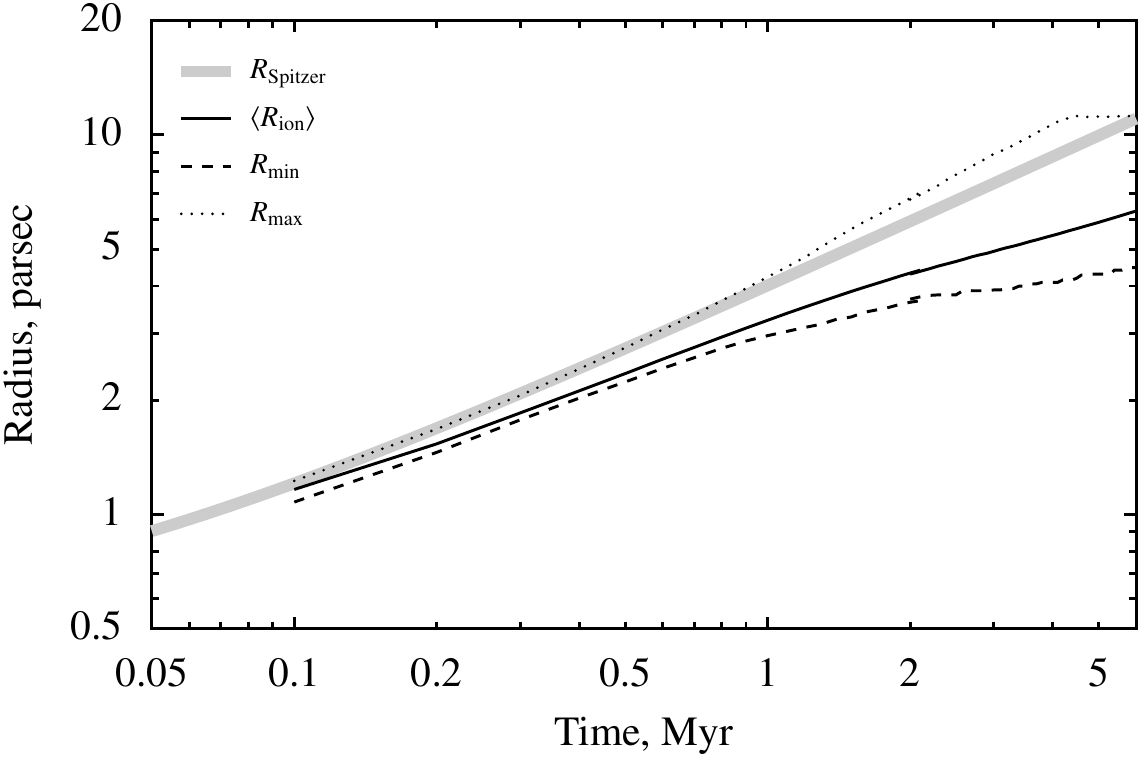}
  \caption{Radius against time for the magnetohydrodynamic expansion
    of a photoionised region in a uniform magnetised medium. Numerical
    MHD simulation (black line) and analytical non-magnetised solution
    for classical Spitzer \ion{H}{2} region expansion (gray
    line). Also shown are the polar radius (dotted line) and
    equatorial radius (dashed line). The break in the lines at \(t =
    2\)~Myr is a result of plotting velocities from two simulations at
    different numerical resolutions.}
\label{fig:mhd_rad}
\end{figure}
\begin{figure}
\centering
  \includegraphics[width=\linewidth]{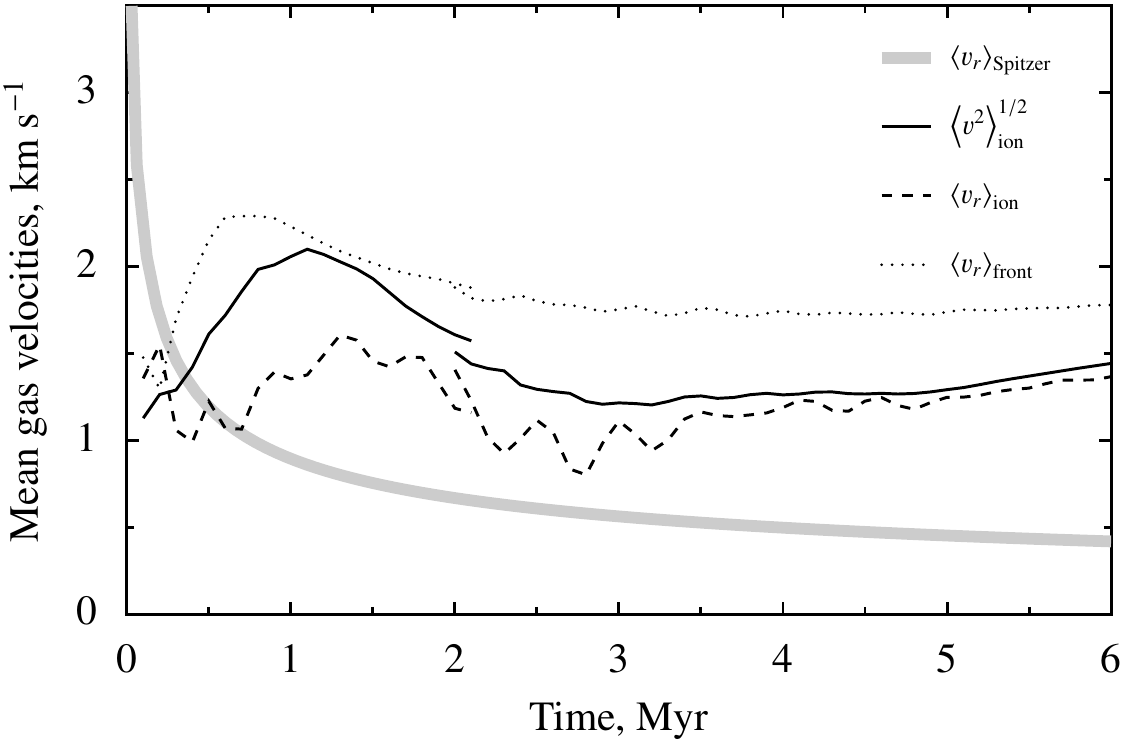}
  \caption{Mean velocities against time for the magnetohydrodynamic
    expansion of a photoionised region in a uniform magnetised
    medium. Rms velocity (black line) and analytical
    non-magnetised solution for classical Spitzer \ion{H}{2} region
    expansion (thick grey line). Also shown are the mean radial
    velocity in the ionised gas (dashed line) and the velocity of the
    gas at the ionization front (dotted line). The break in the lines
    at \(t = 2\)~Myr is a result of plotting velocities from two
    simulations at different numerical resolutions. }
\label{fig:mhd_vel}
\end{figure}
Finally, in Figs.~\ref{fig:mhd_rad} and \ref{fig:mhd_vel}, we
examine the expansion of the magnetised \ion{H}{2}
region. Fig.~\ref{fig:mhd_rad} shows the evolution of the radius of
the photoionised region with time. The mean radius is consistently
below the classical Spitzer expansion law, unlike the non-magnetic
case (see Fig.~\ref{fig:nomhd_rad}). In the direction perpendicular to
the magnetic field, the expansion slows with time, as shown by the
evolution of the minimum radius. In the direction along the field
lines, the expansion initially ($t \leq 1$~Myr) follows the classical
Spitzer law but then accelerates as material is channelled along the
field lines and out of the poles of the \ion{H}{2} region. This
expansion abruptly stops just after 4~Myr, which corresponds to the
time when the ionisation front can no longer expand along the field
lines because the flow has become one dimensional.

In Fig.~\ref{fig:mhd_vel} the mean radial and root-mean-square (rms)
velocities of the ionised gas are plotted together with the analytical
value and also the mean velocity of the ionisation front. The
discontinuity at 2~Myr is due to the change in resolution of the
simulation. The calculated values are all above the analytical values
and indicate that typical velocities in the photoionised gas are of
the order 1--2~km~s$^{-1}$ throughout the simulated lifetime of the
\ion{H}{2} region. The mean radial velocity shows the same sort of
ringing that we saw in the non-magnetised simulation (see
Fig.~\ref{fig:nomhd_vel}). After about 3~Myr, the gas velocities
become dominated by the one-dimensional flow being channelled along
the field lines.

\section{\hii{} Region Expansion in a Magnetised Turbulent Medium}
\label{sec:turbhii}
In this section we describe the results of radiation MHD simulations
of \ion{H}{2} region expansion in a more realistic medium where
neither the initial density distribution nor magnetic field are
uniform. We consider both an O star ionising flux and one more
appropriate to a B star in order to study the effect of the magnetic
field on the ionised gas and on the neutral gas of the PDR.

\subsection{Initial Set Up}
\label{sec:setup}
\NEW{As initial conditions we take the results of a $256^3$
driven-turbulence, ideal magnetohydrodynamic simulation by
\citet{2005ApJ...618..344V}, specifically the simulation with initial isothermal Mach number $M_\mathrm{s} = 10$, Jeans number $J = 4$ and plasma beta $\beta = 0.1$. The original \citet{2005ApJ...618..344V}
simulations are scale free and are characterised by the three
nondimensional quantities $M_\mathrm{s} = \sigma/c$ (the rms sonic
Mach number of the turbulent velocity dispersion $\sigma$),
$J\equiv L/L_\mathrm{J}$ (the Jeans number, giving the size of the
computational box $L$ in units of the Jeans length $L_\mathrm{J}$), and
the plasma beta, $\beta \equiv 8\pi \rho_0 c^2/B_0^2$ (ratio of
thermal to magnetic pressures).}

\NEW{We set the scaling by choosing the computational
box size to be 4~pc, giving the ambient temperature as $T_0
\sim 5$~K, the mean atomic number density as $\langle n_0 \rangle =
1000$~cm$^{-3}$ and the mean magnetic field strength as $B_0 = 14.2
\mu$G.  We choose as our starting point the time in the evolution when
there is one collapsing object. At this point, the mean values of the
magnetic field the number density are unchanged, since flux and mass
are conserved, while the mean plasma beta is now $\beta = 0.032$,
which is consistent with the rms value of the magnetic field $B_{\rm
rms} = 24.16\, \mu$G.} That is, the turbulence has enhanced the rms
magnetic field by a factor $\eta = 24.16/14.2$, which leads to a
change by a factor $\eta^2 \sim 3$ in the value of $\beta$ over the
original, uniform conditions.

As in our previous paper \citep{2006ApJ...647..397M}, we place our
ionising source in the centre of the collapsing object, remove $32
M_{\odot}$ of material (corresponding to the mass of the star formed)
and take advantage of the periodic boundary conditions of the
turbulence simulation to move the source to the centre of the grid. We
also subtract the source velocity from the whole grid. We consider two
ionising sources: one corresponds approximately to an O9 star
\citep{1996ApJ...460..914V}, having an effective temperature $T_\mathrm{eff} = 37500$~K and an ionising photon rate of $Q_\mathrm{H} =
10^{48.5}$~photons~s$^{-1}$, while the other ionising source has $Q_\mathrm{H} = 5
\times 10^{46}$~photons~s$^{-1}$, which corresponds to B0.5.

\subsection{Results} 
\label{sec:results}
In this section we present our results both for the O star and the B
star and in each case for the magnetic and also for the purely
hydrodynamic simulations. In this way we can assess the importance of
the magnetic field and the ionising photon flux on the evolving
structures in the expanding \ion{H}{2} regions and surrounding neutral
gas. We first present emission-line images of the simulated \ion{H}{2}
regions for both O and B stars in the magnetic and pure hydrodynamic
cases when they have reached a similar size, which corresponds to
200,000~yrs of evolution for the O star case and $10^6$~yrs of
evolution in the B star case. This permits a direct comparison with
observations at optical and longer wavelengths. We then examine the
expansion and time evolution of the ionised, neutral and molecular
components using global statistical properties. The importance of the
magnetic field in the different components is then studied for the
particular O and B star simulations presented earlier. 

Although the ionised/neutral transition of hydrogen is treated
explicitly in a detailed manner in our radiation-MHD code, the
neutral/molecular transition is treated much more approximately. We
simply assume that the molecular fraction is a predetermined function
of the extinction $A_V$ from the central star to each point in our
simulation: $n_{\mathrm{mol}}/n_{\mathrm{neut}} = \left(1 +
e^{-4(A_V - 3)} \right)^{-1}$, which was determined by fitting to
the Cloudy models described in the Appendix of
\citet{2009MNRAS.398..157H}.

A rough estimate of the importance of the magnetic field in our
turbulent media simulations can be obtained by calculating the
critical radius and timescale, $R_\mathrm{m}$ and $t_\mathrm{m}$ (see
Eqs.~\ref{eq:rm} and \ref{eq:tm}), for the initial value of the
magnetic field and the mean density. Using $B_\mathrm{rms} = 24.16\,\mu$G
and $\langle n \rangle = 10^3$~cm$^{-3}$, and assuming a mean particle
mass of 1.3~m$_\mathrm{H}$, the initial representative Alfv\'en speed in the
neutral gas is $v_\mathrm{A} \equiv B_\mathrm{rms}/(4\pi \rho)^{1/2} \simeq
1.46$~km~s$^{-1}$. In the photoionised gas, the sound speed is $c_\mathrm{i} =
9.8$~km~s$^{-1}$. For the strong ionising source (O star), the Str\"omgren
radius in a uniform medium of density equal to the mean density would
be $R_\mathrm{0} = 0.45$~pc, giving a critical radius $R_\mathrm{m} =
5.7$~pc and time $t_\mathrm{m} = 2.2$~Myr for when the magnetic field
starts to become important. If, instead of the rms magnetic field
we use the mean magnetic field, $\langle B \rangle = 14.2\,\mu$G, then
the corresponding values are $R_\mathrm{m} = 11.86$~pc and
$t_\mathrm{m} = 7.9$~Myr. For the weak ionising source (B star), the numbers
are $R_\mathrm{0} = 0.113$~pc, $R_\mathrm{m} = 1.43$~pc
($R_\mathrm{m} = 2.98$~pc) and $t_\mathrm{m} = 0.55$~Myr
($t_\mathrm{m} = 1.98$~Myr) depending on whether the rms or mean
magnetic field is considered. 

Thus, for the strong ionising source, the critical diameter is larger
than the dimension of the computational box, hence we do not expect
that the magnetic field would have much effect on the expansion of the
\ion{H}{2} region in this case. However, for the weak ionising source,
it is not clear whether the rms or the mean value of the magnetic
field is most important for determining the global evolution of the
\ion{H}{2} region. If the rms field is the determining parameter
then the fact that the critical diameter in this case is smaller than
the size of the computational box leads us to expect that the magnetic
field will affect the global expansion of the \ion{H}{2} region. On the other
hand, if it turns out that the mean magnetic field is most important,
then the effect will not be noticed within our computational size and
timescales. Furthermore, given the non-uniform initial conditions, the
timescale for the magnetic field to become important could vary
depending on the local gas sound speed to Alfv\'en speed ratio.

\subsubsection{Morphologies and Images}
\label{sec:morph}
\begin{figure*}
\includegraphics[width=\linewidth]{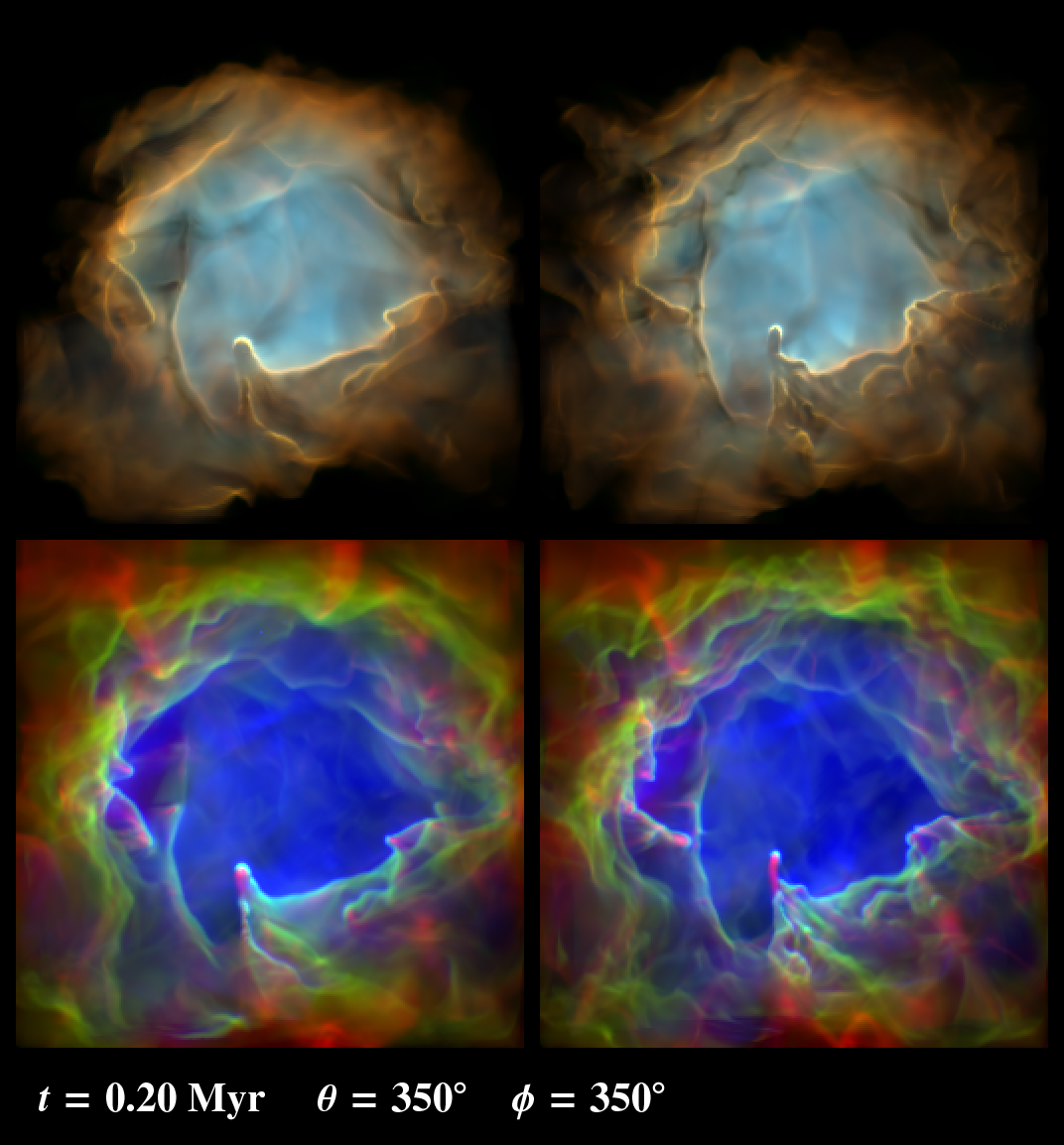}
\caption{Simulated optical (top) and long-wavelength (bottom) emission
   for an \protect{\ion{H}{2}} region around an O star after
  200,000~yrs of evolution for MHD (left) and purely hydrodynamic
   (right) simulations. \NEW{The viewing angle is \(\theta =
    350^\circ\), \(\phi = 350^\circ\), where \(\theta\) is the polar
    angle, measured from the \(x\)-axis, and \(\phi\) is the azimuthal
    angle, measured around the \(x\)-axis.} \textit{Top}: Synthetic narrowband optical emission-line images in the
  light of [\protect{\ion{N}{2}}] 6584~\AA (\textit{red}), H$\alpha$ 6563~\AA
  (\textit{green}), and \ion{O}{3} 5007~\AA
  (\textit{blue}). \textit{Bottom}: Synthetic images in the
  light of 6\,cm radio free-free emission (\textit{blue}), generic PAH
  (\textit{green}) and molecular gas column density (\textit{red})}
\label{fig:images_O}
\end{figure*}
\begin{figure*}
\includegraphics[width=\linewidth]{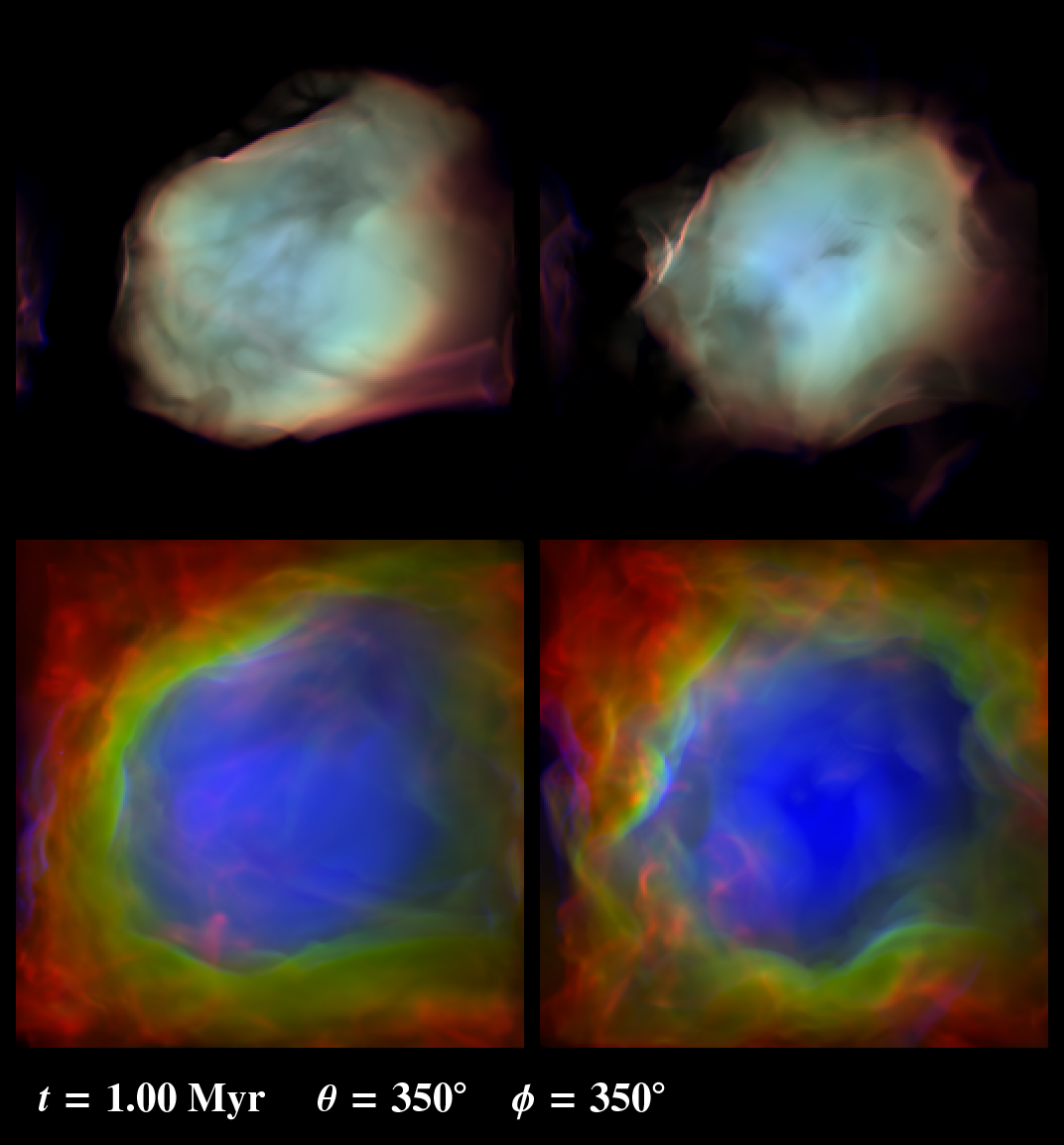}
\caption{Simulated optical (top) and long-wavelength (bottom) emission
  for an \protect{\ion{H}{2}} region around a B star after $10^6$~yrs
  of evolution for MHD (left) and purely hydrodynamic (right)
  simulations. \NEW{The angles \(\theta\) and \(\phi\) are defined in
    the caption to Fig.~\ref{fig:images_O}.}  \textit{Top}: Synthetic
  narrowband optical emission-line images in the light of
  [\protect{\ion{N}{2}}] (\textit{blue}), H$\alpha$ (\textit{green}),
  and \ion{S}{2} (\textit{red}). \textit{Bottom}: Synthetic images in
  the light of 6\,cm radio free-free emission (\textit{blue}), generic
  PAH (\textit{green}) and molecular gas column density
  (\textit{red})}
\label{fig:images_B}
\end{figure*}

We start our comparison by considering the morphological appearance of
the \ion{H}{2} regions and the surrounding neutral and molecular
material.  In Figs.~\ref{fig:images_O} and \ref{fig:images_B} we show
two types of image data, one showing mostly the ionised gas, where the
different colours represent emission from optical emission lines, the
other shows synthetic, long-wavelength (infrared and radio) emission,
emphasizing the neutral and molecular material.

\NEW{The optical emission was calculated as described in our previous
  papers \citep{2006ApJ...647..397M, 2009MNRAS.398..157H}, assuming
  heavy element ionisation fractions that are fixed functions of the
  hydrogen ionisation fractions. For the B star models, these
  functions were recalibrated for the softer stellar ionising spectrum
  using the Cloudy plasma code \citep{1998PASP..110..761F}. For the O
  star models, we employ the ``classical'' \textit{HST} red-green-blue
  colour scheme of [\ion{N}{2}], H\(\alpha\), [\ion{O}{3}]. Since the
  [\ion{O}{3}] emission from the B star models is predicted to be very
  weak, in this case we use the scheme [\ion{S}{2}], [\ion{N}{2}],
  H\(\alpha\). In both schemes, the progression from red through green
  to blue corresponds to increasing degree of ionisation inside the
  \ion{H}{2} region. The emission from all the optical lines is
  negligible in the dense neutral/molecular zones, but the dust
  absorption associated with these dense regions is visible in the
  images. Scattering by dust is not included.}
 
\NEW{For the long-wavelength images the emission bands were chosen to
  give a more global view of the simulations than can be provided by
  the optical emission lines. The red band shows the total column
  density of neutral/molecular gas, which very crudely approximates
  far-infrared or sub-mm continuum emission from cool dust. The green
  band of the long-wavelength images shows a simple approximation to
  the mid-infrared emission from polycyclic aromatic hydrocarbons
  (PAH). The PAHs are assumed to reprocess a fixed fraction of the
  local far-ultraviolet radiation field. Detailed calculations
  \citep{2001ApJ...556..501B} show that this assumption is correct to
  within a factor of about two for most PAH emission features over a
  broad range of conditions. The local PAH density is assumed to be
  proportional to the neutral plus molecular hydrogen density, since
  there is strong evidence that PAHs are destroyed in ionised gas
  \citep{1989ApJ...344..791B, 2007ApJ...665..390L}. No attempt is made
  to discriminate between different PAH charge states. The blue band
  of the long-wavelength images shows the 6~cm radio continuum
  emission due to bremsstrahlung in the ionised gas. Apart from a
  slightly different temperature dependence this is very similar to
  the H\(\alpha\) emission, except that it does not suffer any dust
  absorption. 
}

Starting with the O-star case, we see that the morphological
appearance shows the typical characteristics of \ion{H}{2} regions in
turbulent media found in earlier work
\citep{2006ApJ...647..397M,2007MNRAS.377..535D,{2007ApJ...668..980M}},
namely a fairly irregular structure with fingers or pillars, as well
as bar-like features at the edge of the \ion{H}{2} region. Note how
the most prominent finger in the bottom edge of the \ion{H}{2} region
does not exactly point towards the source of ionising photons. In
appearance the simulated \ion{H}{2} regions look strikingly like
observed \ion{H}{2} regions.

Comparing the magnetic with the non-magnetic case, one sees an overall
agreement in the structures; the presence of a magnetic field does not
have a large impact on the global morphology of the \ion{H}{2} region. This
is to be expected as for the O-star case the critical time (see
Eq.~\ref{eq:tm}) has not been reached by the time most of the volume
has become ionised. However, the magnetic field does have an effect on
small scale features in the \ion{H}{2} region. In the magnetic case
the fingers and bars look smoother and broader, due to the magnetic
pressure, which modifies the radiation-driven implosion of these
structures \citep{2009MNRAS.398..157H,{2005Ap&SS.298..183R}}. This
behaviour is visible in both optical and long-wavelength images, but
most clearly in the latter. In section~\ref{sec:discussion} we discuss
the behaviour of the globules in greater depth.

For the B-star case, the critical time does occur within the
simulation.  However, as is apparent from Fig.~\ref{fig:images_B} this
does not lead to a significant deformation of the \ion{H}{2}
region. The reason is that magnetic field itself also has a turbulent
structure, and there is no large-scale field which can impress its
direction on the shape of the \ion{H}{2} region, as in the test case
presented above (see \S~\ref{sec:krum}). Although the magnetic and
non-magnetic results do differ, the overall impression is that they are
still quite similar. In fact, the difference between the O-star and
the B-star cases is much larger than between the magnetic and
non-magnetic cases. For the B-star case, no pillars are found and the
ionised region appears more spherical. The edge of the \ion{H}{2}
region is often more fuzzy, although some sharper bar-like features
are present. The long wavelength images show the reason for this: the
\ion{H}{2} region is embedded in a thick PDR region (with an extent of
$\sim 30\%$ of the radius of the \ion{H}{2} region) which erases much
of the small-scale density fluctuations present in the original cold
molecular medium. Close inspection reveals that the high-density
regions already inside the PDR region photo-evaporate and thus lose
their large density contrast. In the O-star case these are the features which
develop into pillars.  The fact that the temperature in the PDR is
closer to 100~K than to 10~K in combination with the low ionising flux
of the B-star should generally be less conducive to the formation of
pillars, as shown by \citet{2010ApJ...723..971G}.

Careful examination of the images shows that otherwise the effect of
the magnetic field is quite similar to that in the O-star case:
small-scale structures are smoothed out in the magnetised
case. However, the effect is quite marginal and does not give a clear
observable diagnostic for the presence of magnetic fields or not.

\subsubsection{Global Properties of the \protect{\ion{H}{2}} Region
  Evolution}
\label{sec:glob-prop-prot} 
\begin{figure}
\includegraphics[width=\linewidth]{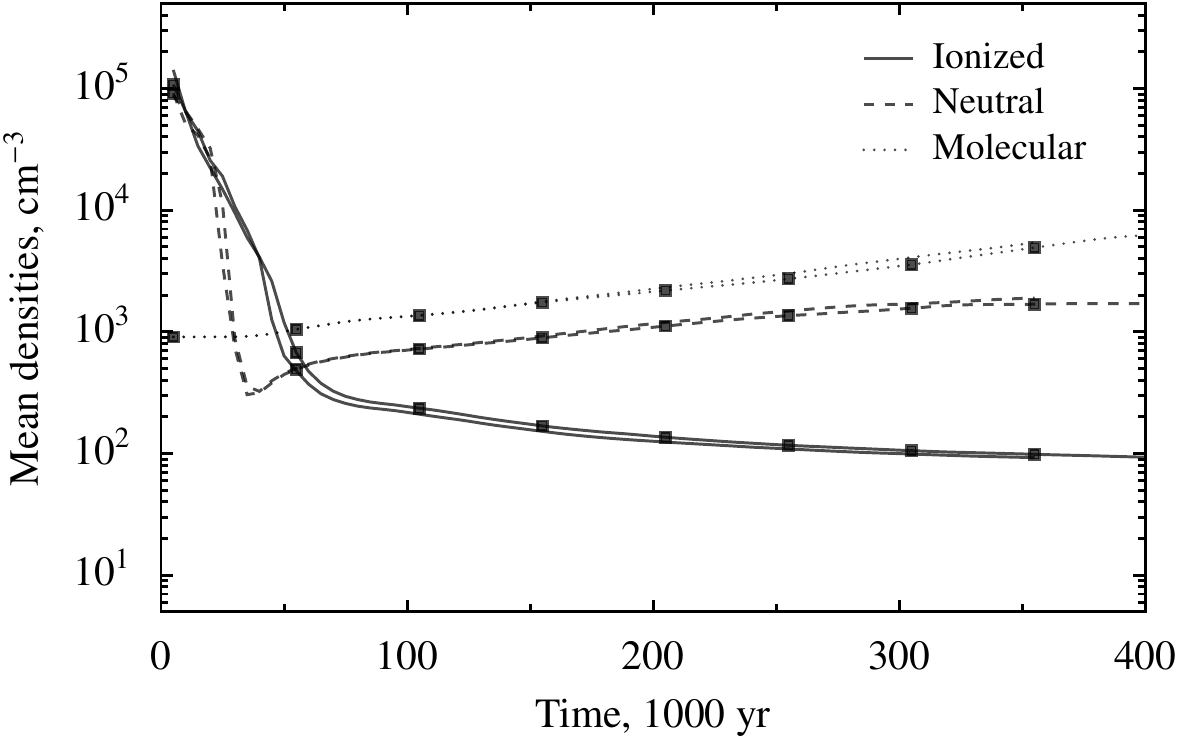}
\caption{Mean densities of the ionised (solid line), neutral (dashed
  line) and molecular (dotted line) components for the evolving
  \protect{\ion{H}{2}} region around the O star. Lines with symbols
  are for the MHD simulation, while lines without symbols are for the
  purely hydrodynamic case.} 
\label{fig:comp1_O}
\end{figure}
\begin{figure}
\includegraphics[width=\linewidth]{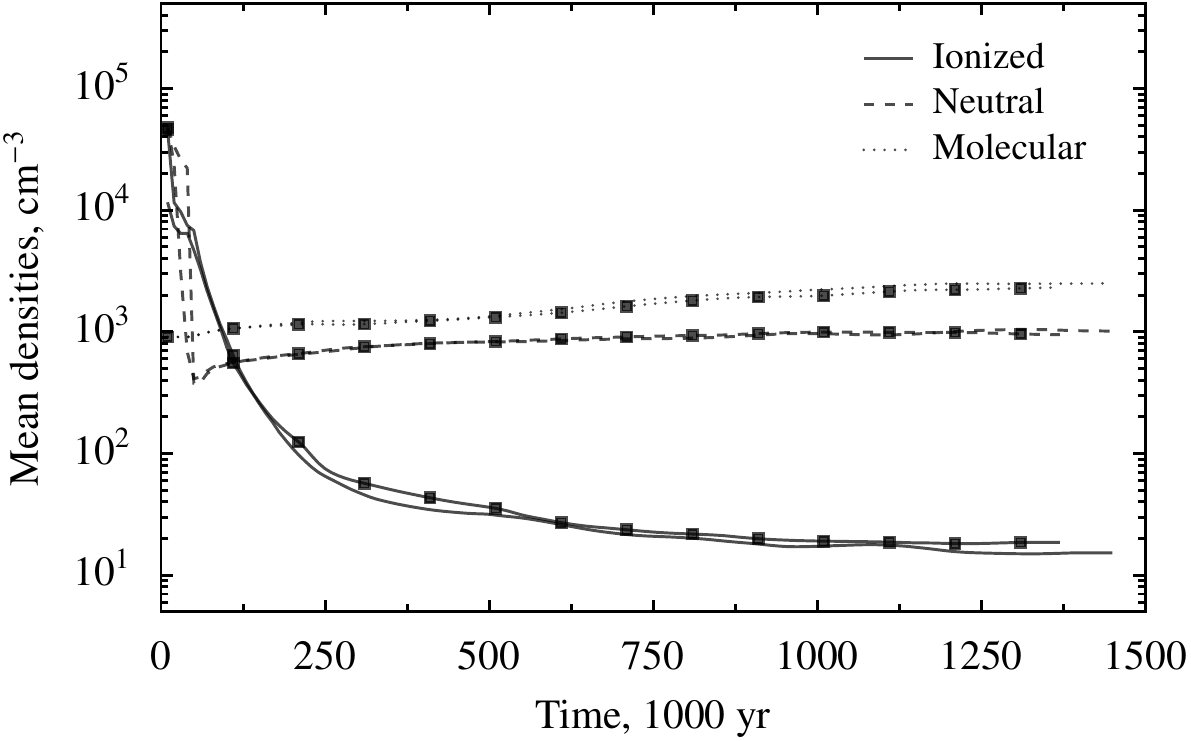}
\caption{Same as Fig.~\protect\ref{fig:comp1_O} but for the B star.}
\label{fig:comp1_B}
\end{figure}

The initial mean density in the computational box is $n_0 =
1000$~cm$^{-3}$ but the material is distributed inhomogeneously with
the densest clumps having densities $> 10^5$~cm$^{-3}$. In
Figs.~\ref{fig:comp1_O} and \ref{fig:comp1_B} we can see how the
mean densities in the ionised, neutral and molecular components evolve
with time and the growth of the \ion{H}{2} regions around the O and B
stars, respectively. In these figures, lines with symbols are for the
MHD simulations and lines without symbols are the purely hydrodynamic
case. We can see immediately that the magnetic field has a negligible
effect on the densities of the different components in both the O and
B star cases. As the evolution of the \ion{H}{2} region proceeds, the
mean density of the molecular gas in the O star simulation increases
slowly with time because the lower density molecular gas becomes ionised or
incorporated into the neutral PDR and also because the dense clumps and
filaments at the edge of the \ion{H}{2} region are compressed due to
radiation-driven implosions, thereby increasing their density. By the end of
the simulation, only these densest clumps remain, since the majority
of the computational box has been ionised. In the B star case, the
mean density of the molecular gas remains roughly constant throughout
the simulation. This is because the \ion{H}{2} region does not break
out into regions of lower density and remains confined within the
molecular gas for the duration of the simulation. Also, there is less
fragmentation in the B star simulation because density inhomogeneities
in the neutral region of the PDR are smoothed out by photoevaporation
flows due to FUV radiation before the ionisation front reaches them
(see Fig.~\ref{fig:images_B}).

By the end of the respective simulations, the mean density in the
ionised gas around the O star is $\sim 100$~cm$^{-3}$, while
that around the B star is $\sim 10$~cm$^{-3}$ even though the spatial
extent is much smaller. This is because the latter is a much weaker ionising
source. In both cases, the density of the neutral PDR gas tends to a
roughly constant value of $10^3$~cm$^{-3}$. 

\begin{figure}
\includegraphics[width=\linewidth]{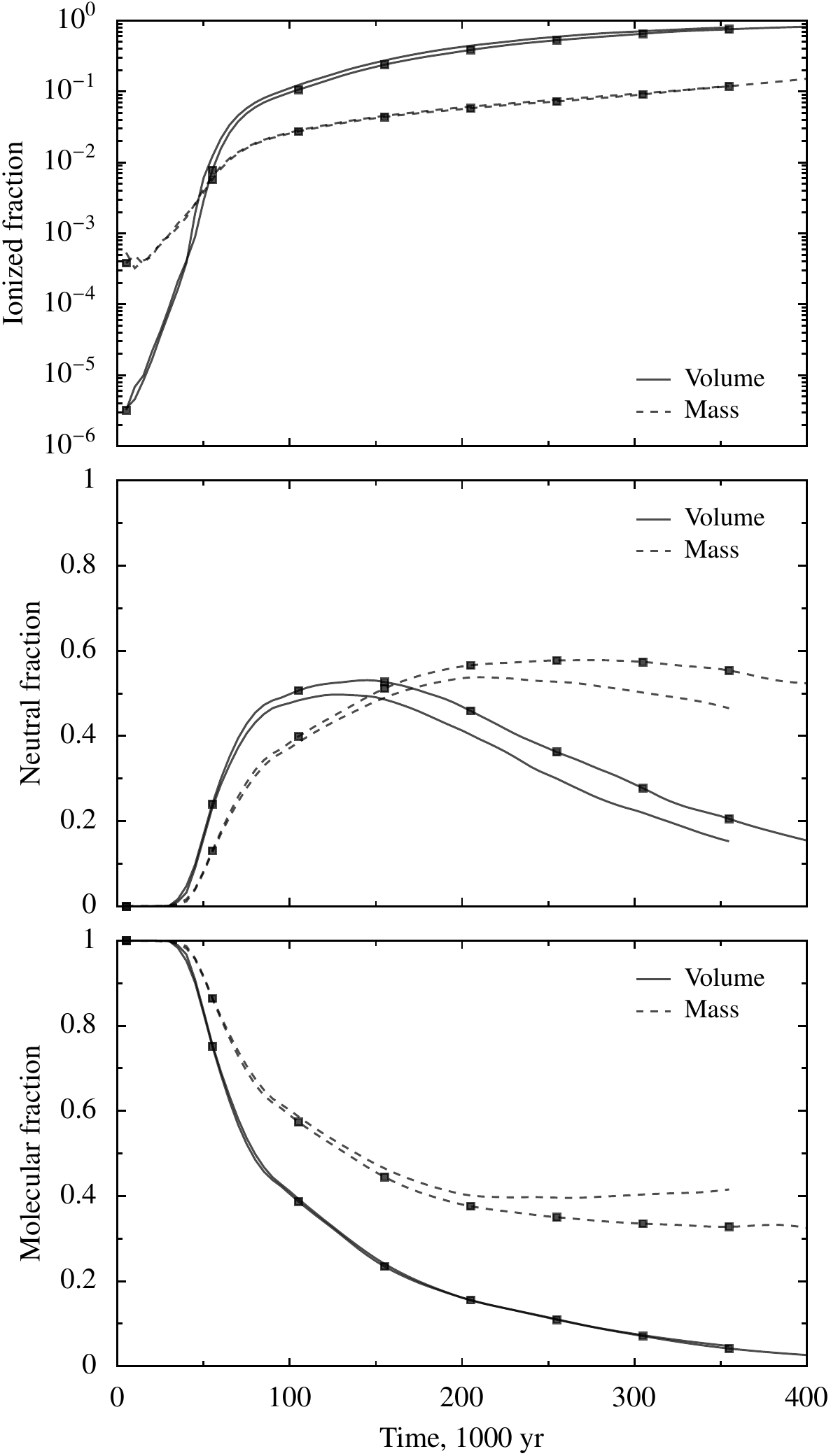}
\caption{Ionised, neutral and molecular gas fractions by volume (solid
  lines) and mass (dashed lines)  for the evolving
  \protect{\ion{H}{2}} region around the O star. Lines with symbols
  are for the MHD simulation, while lines without symbols are for the
  purely hydrodynamic case.} 
\label{fig:comp2_O}
\end{figure}
\begin{figure}
\includegraphics[width=\linewidth]{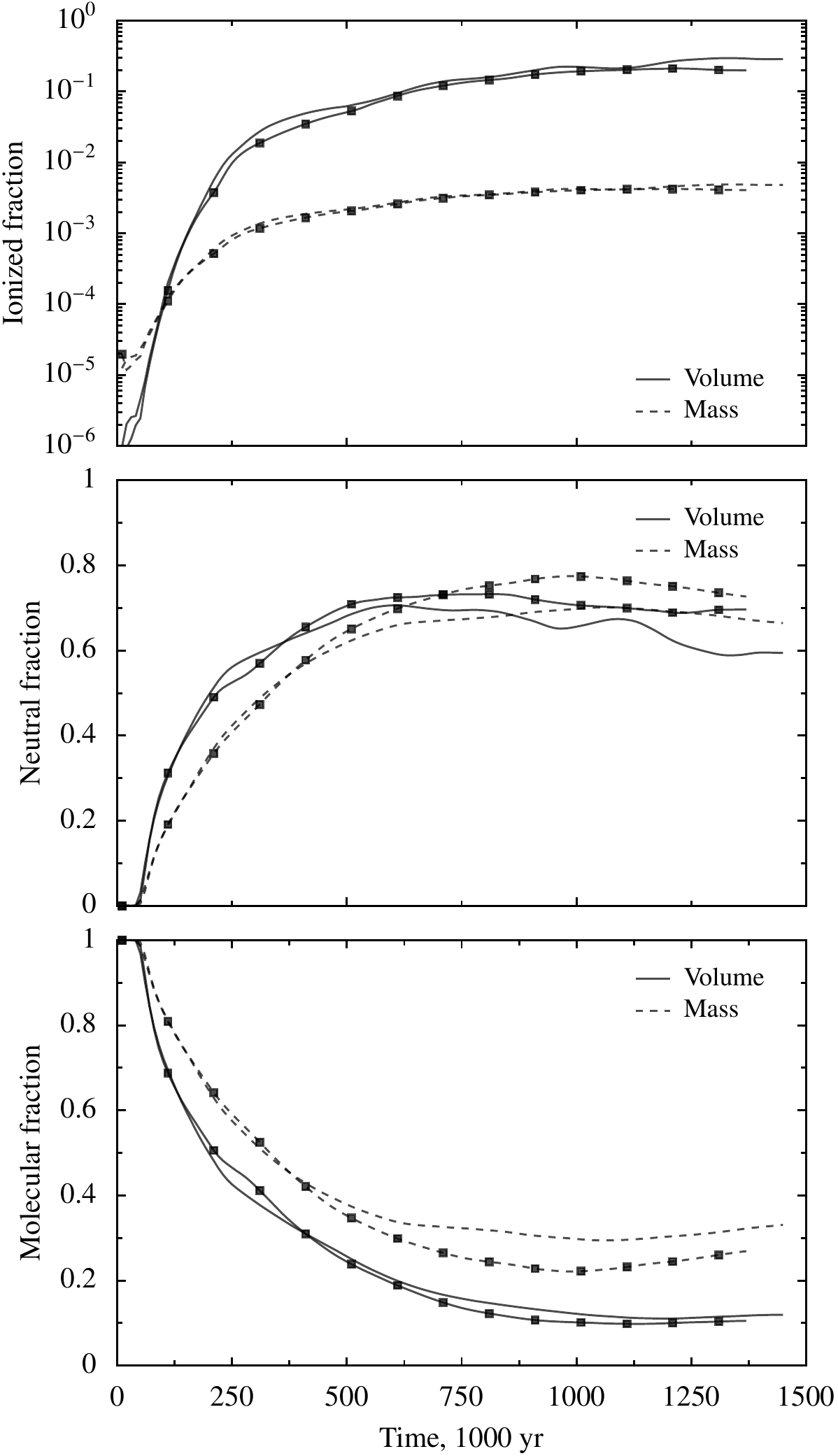}
\caption{Same as Fig.~\protect\ref{fig:comp2_O} but for the B star.}
\label{fig:comp2_B}
\end{figure}

In Figs.~\ref{fig:comp2_O} and \ref{fig:comp2_B} we compare how the
ionised, neutral and molecular gas component fractions vary throughout
the evolution of the \ion{H}{2} regions. We calculate both volume and
mass fractions for each component. The ionised gas in both the O and B
star cases is thermally dominated and there is no discernible
distinction between the MHD and purely hydrodynamic results. In the O
star case, the ionised fraction comes to occupy 90\% of the volume but
only 10\% of the mass of the computational box by the end of the
simulation. In the B star case, the \ion{H}{2} region does not manage
to globally break out of the computational box and so even at the end
of the simulation the volume occupied by ionised gas is less than
30\%. The mass fraction is $< 1\%$ in this case, since the density in
the ionised gas is low. The neutral gas is most affected by the
magnetic field for both O and B star
simulations. Figs.~\ref{fig:comp2_O} and \ref{fig:comp2_B} show that
the MHD simulation has both higher mass and higher volume fractions of
neutral gas than the purely hydrodynamic results. The neutral
component is more important in the B star case than in the O star
case, reaching $\sim 70\%$ of the mass and volume fractions after
500,000~yrs and remaining roughly constant thereafter. In the O star
case, however, the mass fraction remains roughly constant for the
second half of the simulation while the volume fraction decreases
sharply. This difference in behaviour can be attributed to the greater
fragmentation in the O star case (both MHD and HD) compared to the B
star. The neutral material in the B star simulation is distributed in
a quite broad, smooth, almost uniform density region around the
\ion{H}{2} region, whereas in the O star simulation the neutral region
around the photoionised gas is relatively thin but there is also
neutral material inside the dense clumps and filaments that are formed
by radiation-driven implosion, which have high density but low
volume. 

\begin{figure}
\includegraphics[width=\linewidth]{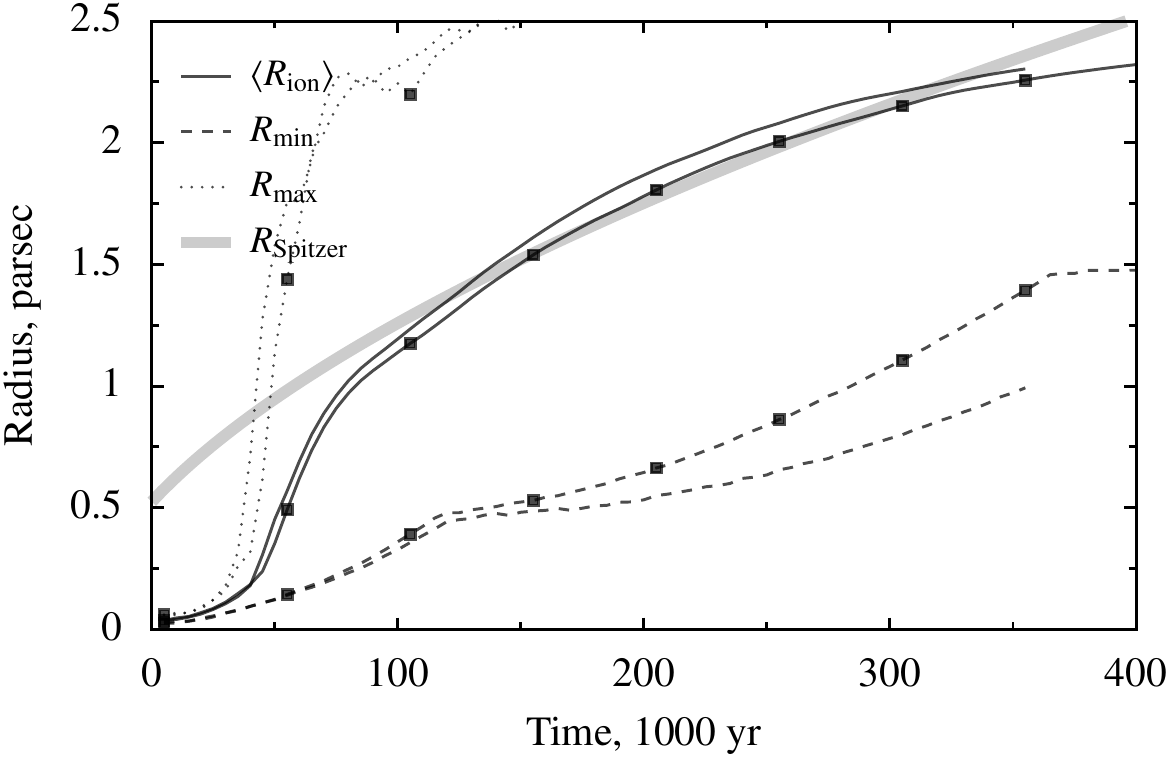}
\caption{Expansion of the \protect{\ion{H}{2}} region with time for
  the O star. The thick grey line represents the analytical
  solution. Also shown are the mean radius (solid lines), the maximum
  radius (dotted lines) and the minimum radius (dashed lines) of the
  ionisation front. Lines with symbols 
  are for the MHD simulation, while lines without symbols are for the
  purely hydrodynamic case.} 
\label{fig:comp2b_O}
\end{figure}
\begin{figure}
\includegraphics[width=\linewidth]{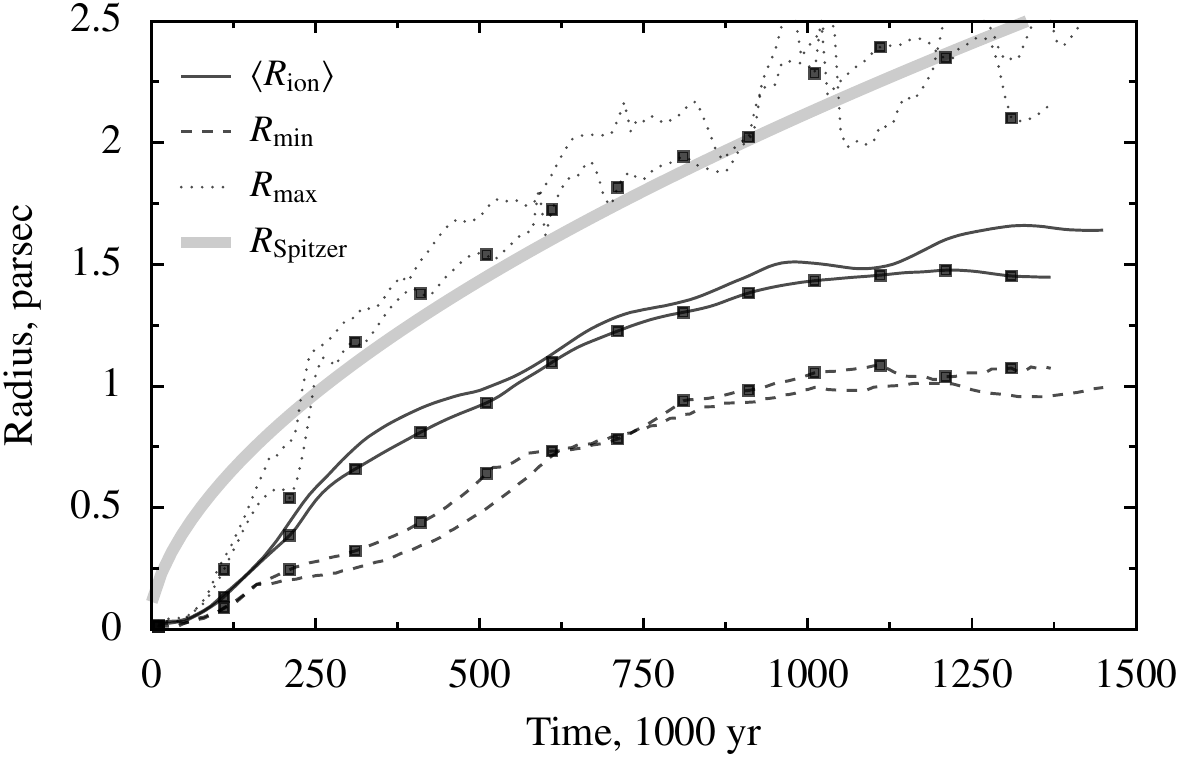}
\caption{Same as Fig.~\protect\ref{fig:comp2b_O} but for the B star.}
\label{fig:comp2b_B}
\end{figure}
The expansion of an \ion{H}{2} region in a clumpy medium is not so
easy to characterise. In Figs.~\ref{fig:comp2b_O} and
\ref{fig:comp2b_B} we plot a variety of different measures of the
radius of the photoionised region as functions of time, together with
the analytical solution obtained using Equation~\ref{eq:spitz} for a
medium of uniform density $n_0 = 10^3$~cm$^{-3}$. We calculate the
mean radius\footnote{We define the mean radius as the radius of a
sphere with the same volume as the real \ion{H}{2} region.} and also
the minimum and maximum radii of the ionisation front. For the
\ion{H}{2} region around the O star, the mean radius closely follows
the analytical solution for a long period of the simulation, possibly
because the filling factor of the dense clumps and filaments which
retard the ionisation front is small. The maximum radius for the
ionisation front for the O star simulation quickly becomes equal to
the distance to the edge of the computational box and therefore has no
physical meaning after this point. Until it leaves the box, the
maximum radius represents the direction in which the ionisation front
is able to expand most quickly, i.e. a path of lowest density radially
outward from the star.

For the B star simulation, the mean radius is always below the
analytical solution. Since this is true for both the MHD and purely
hydrodynamic simulations, it cannot be due to the magnetic field. The
reason is that for the B star, the initial clump in which the star is
embedded has a greater effect on the subsequent evolution of the
\ion{H}{2} region than in the O star case. In fact, we see that the
maximum radius of the ionisation front in the B star \ion{H}{2} region
does follow the analytical solution. The maximum radius follows the
expansion of ionisation front in the direction where it was first able
to break out of the dense clump in which the star was embedded. We
have already seen that the neutral medium around the \ion{H}{2} region
in the B star case takes on a smooth and homogeneous appearance and
that the mean molecular and neutral gas densities are roughly constant
in time. Once the ionisation front has managed to break out of the
clump, therefore, it follows the expansion law for a uniform medium in
this mean density. In the majority of directions, however, the
ionisation front takes much longer to break out of the initial clump
and so the mean ionisation front radius is retarded compared to the
analytical solution.

In both O and B star simulations, the minimum radius of the ionisation
front indicates the presence of fingers and clumps of neutral material
within the \ion{H}{2} region. For the O star simulation we have
already mentioned how the magnetic field suppresses fragmentation, and
this is reflected in the graph of minimum radius with time: fingers
and globules formed in the purely hydrodynamic simulation are denser
and survive longer closer to the star compared to the MHD simulation. 
In the B star case there is much less fragmentation and we see no
fingers and clumps of neutral gas within the \ion{H}{2} region and the
minimum radii of both MHD and hydrodynamic simulations seen in
Fig.~\ref{fig:comp2b_B} bear this out.

\begin{figure}
\includegraphics[width=\linewidth]{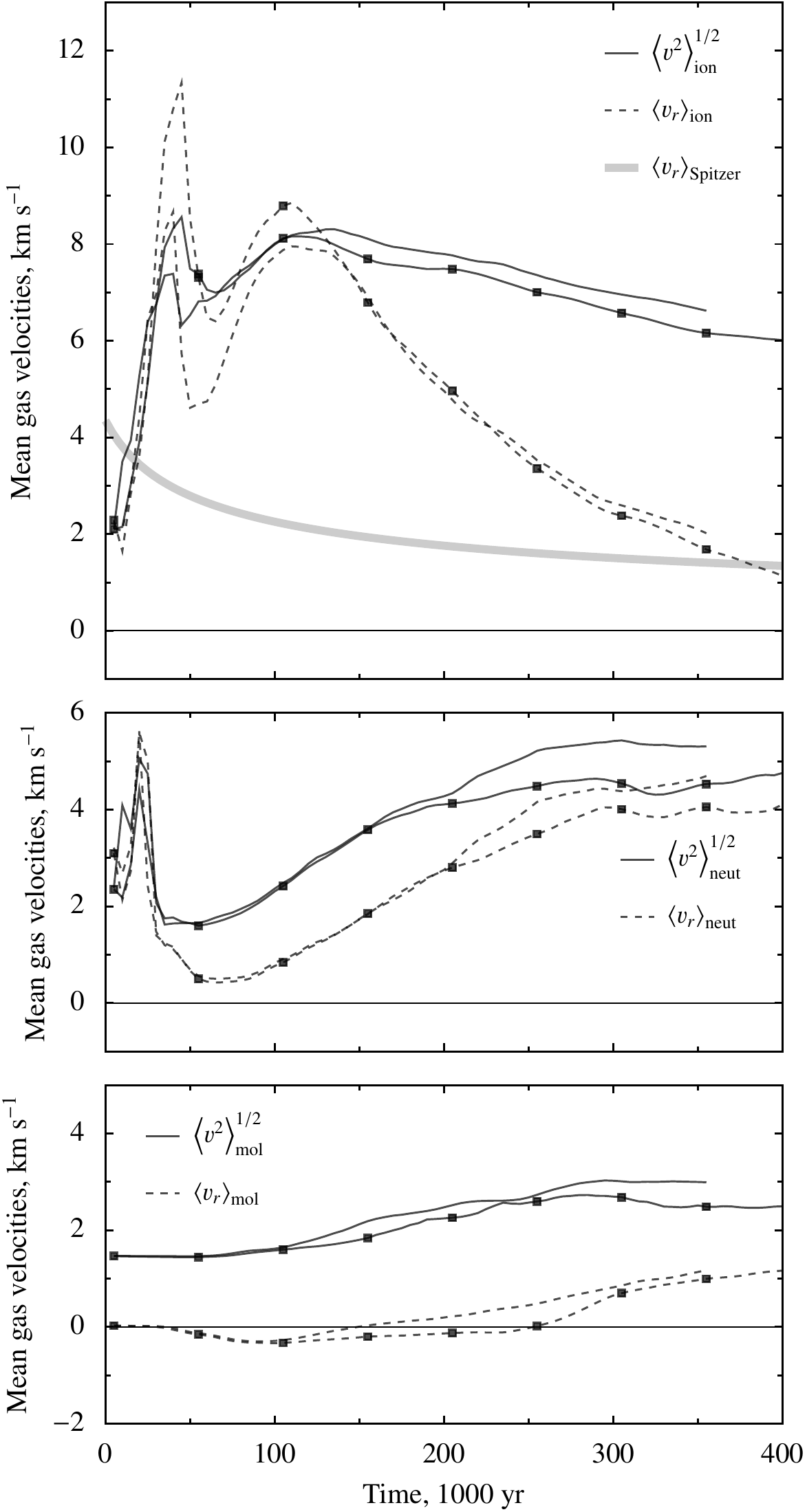}
\caption{Mean gas radial velocities within the ionised, neutral and molecular
  gas components for the O star simulation. The thick grey line represents the analytical
  mean radial velocity. Also shown are the rms radial velocities (solid lines) and the
  mean radial velocity (dashed lines) of the
  gas. Lines with symbols 
  are for the MHD simulation, while lines without symbols are for the
  purely hydrodynamic case.} 
\label{fig:comp3_O}
\end{figure}
\begin{figure}
\includegraphics[width=\linewidth]{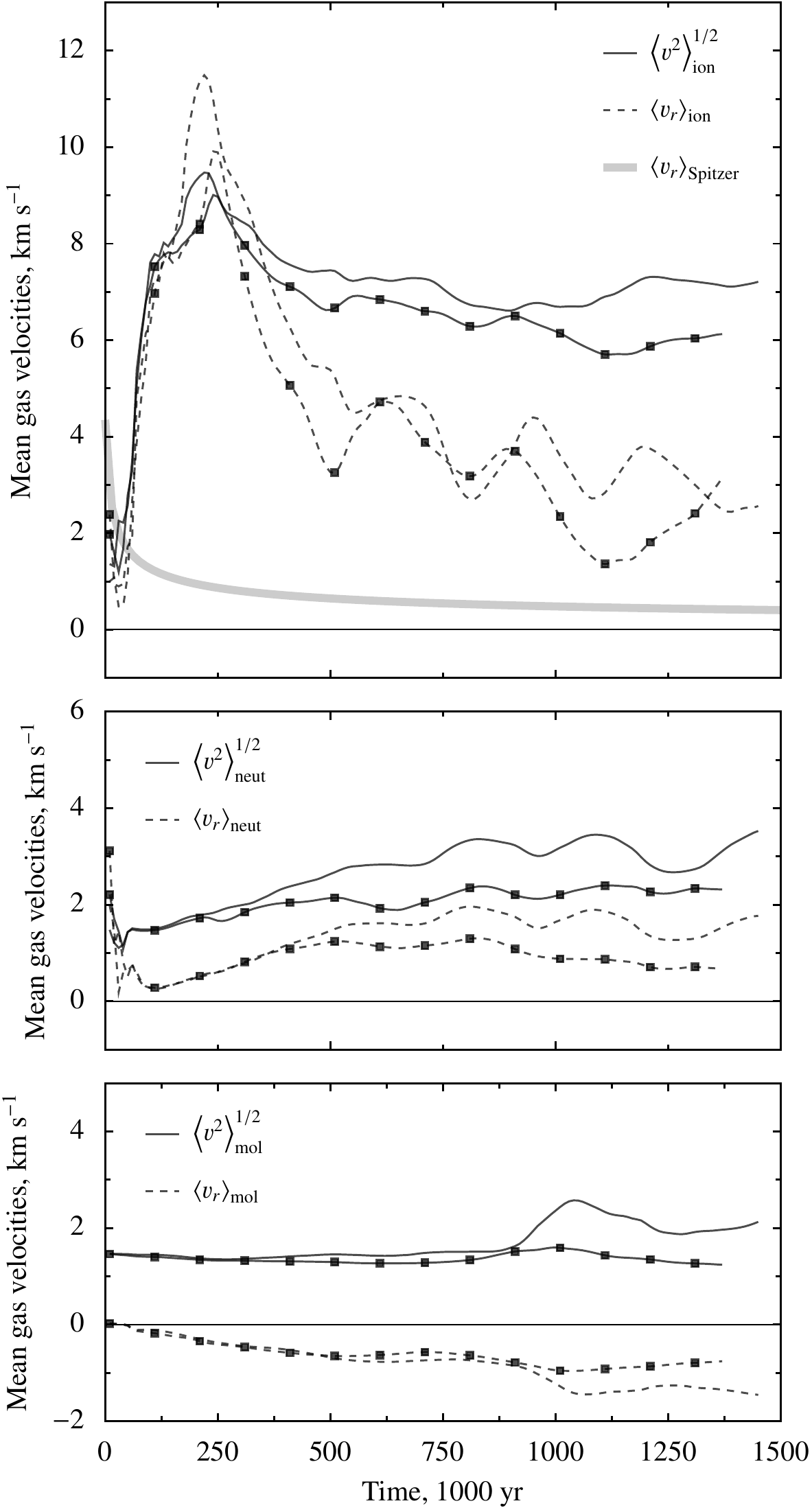}
\caption{Same as Fig.~\protect\ref{fig:comp3_O} but for the B star.}
\label{fig:comp3_B}
\end{figure}

In Figs.~\ref{fig:comp3_O} and \ref{fig:comp3_B} we show the mean
radial velocities for the three gas components. Again, we compare the
results of both MHD (lines with symbols) and purely hydrodynamic
simulations. Beginning with the ionised gas component, we see that for
the O star the rms velocities reach a peak of about 8~km~s$^{-1}$,
and then decreases very slowly, which is what we saw in our previous
paper \citep{2006ApJ...647..397M}. The mean radial velocity, on the
other hand, peaks at 8--10~km~s$^{-1}$ then declines quite
quickly. The rms velocities are due to interacting photoevaporated
flows within the \ion{H}{2} region, which lead to gas flowing inwards
as well as outwards \citep{2003RMxAC..15..175H}. The mean radial
velocity traces the general expansion of the photoionised gas. We see
that for the non-uniform medium, the initial expansive motions are
more rapid than for the simple analytical model but this is just
because the gas first begins to expand into regions of lowest density
\citep{2002ApJ...580..969S}. There is a small difference between the
rms velocities in the MHD and hydrodynamic cases, with the latter
being marginally higher. This could be due to the greater degree of
fragmentation in the hydrodynamic simulations which, in turn, leads to
stronger photoevaporated flows. 

In the B star simulation the rms velocities are similar to the O
star case just discussed, which suggests that photoevaporated flows
are important in this \ion{H}{2} region, too. The pure hydrodynamic
simulation shows slightly higher velocities than the MHD case, which
indicates that the magnetic field plays some role in reducing the
importance of photoevaporated flows within the ionised gas. The mean
radial velocities show the same sort of ringing seen in the test
problem of \ion{H}{2} region expansion in a uniform density medium
(see Fig.~\ref{fig:nomhd_vel}). The period of the oscillations is of
order $2\times10^5$~yrs, which is consistent with the sound crossing
time of the photoionised region. Note that in both O and B star
simulations, the actual mean radial velocities are considerably higher
than the corresponding analytical values for expansion in a uniform
medium with density equal to the mean density of the initial
computational box ($n_0 = 1000$~cm$^{-3}$).

The neutral gas velocities show more separation between hydrodynamic
and MHD results, with the hydrodynamic velocities being consistently
higher in both mean and rms velocities. Interestingly, the
different simulation velocities appear to depart from each other at a
specific point in time, which is different depending on whether the
mean or rms velocity is considered. This is true for both O and B
star simulations. In the O star case, after the initial acceleration,
the neutral gas velocities reach 4--5~km~s$^{-1}$, while for the B
star the velocities are 1--2~km~s$^{-1}$ lower and show evidence for
ringing as for the ionised gas.

The initial molecular gas velocities reflect the Mach 10 supersonic
turbulence of the underlying density distribution. At early times,
there is evidence of residual infalling motions, i.e. negative radial
velocities, again from the initial conditions. These persist longer in
the B star simulations than in the O star simulations. This is because
the O star rapidly ionises out to the edge of the computational box,
and the molecular material which remains is being accelerated radially
away from the central star in clumps that are experiencing the rocket
effect. In the B star case, most of the molecular material surrounding
the \ion{H}{2} region remains undisturbed, and so the imprint of the
initial conditions remains in this gas until the end of the
simulation.

In summary, we find slight differences in the expansion of \ion{H}{2}
regions between MHD and purely hydrodynamic simulations but the
greatest differences are due to the different ionising and FUV fluxes
of the central O or B star. Most of the differences between the MHD
and hydrodynamic cases are due to the lesser amount of fragmentation
in the former, indicating that magnetic fields provide some support to
the neutral gas against radiation-driven implosions.

\subsection{Magnetic Quantities}
\label{sec:magnetic-quantities}
\begin{figure*}
\includegraphics[width=\linewidth]{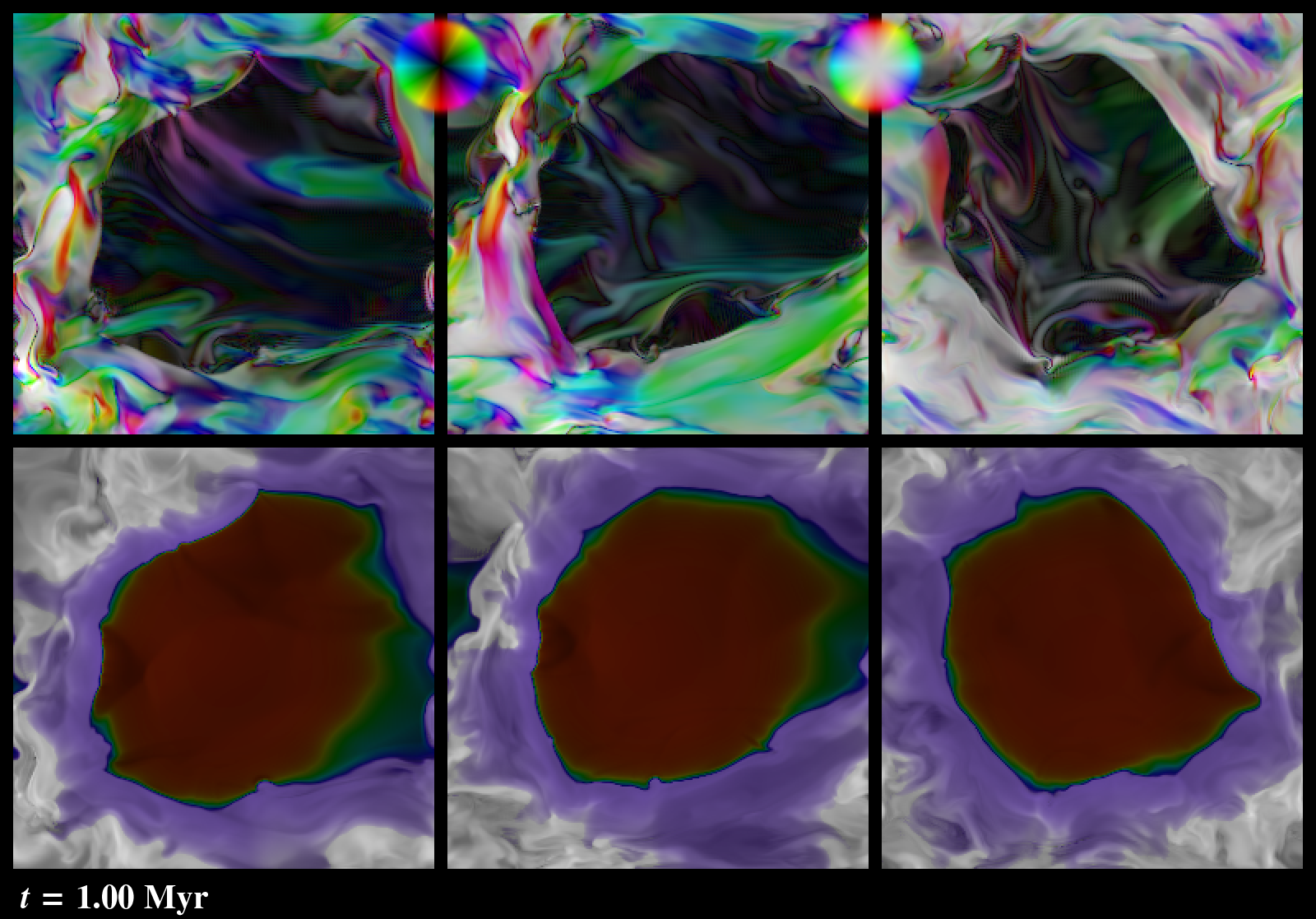}
  \caption{Magnetic field and photoionised gas distribution in the
  central $x$-$y$ (left), $x$-$z$ (centre) and $y$-$z$ planes for the
  B star simulation after $10^6$~yrs. Colours have the same meaning as
  in Fig.~\protect\ref{fig:krum1}.}
\label{fig:cuts_Bstar}
\end{figure*}
\begin{figure}
\includegraphics[width=\linewidth]{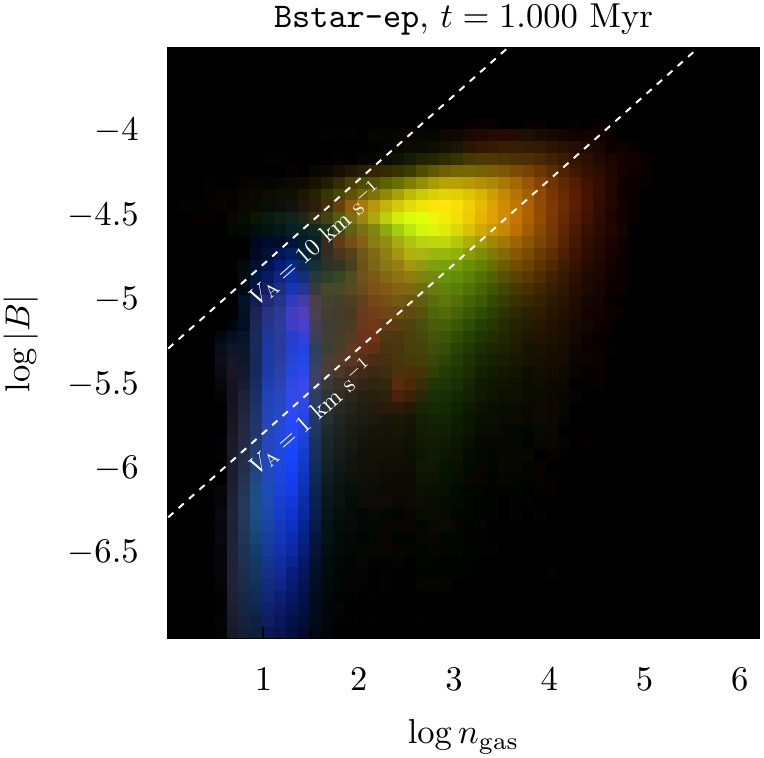}
  \caption{Magnitude of B-field (Gauss) against number density
    (cm$^{-3}$) for volume-weighted joint distributions in the case of
    the B-star at $t=10^6$~years. The diagonal lines correspond to
    constant Alfv\'en speeds of 10 and 1~km~s$^{-1}$. The three colours
    represent ionised (blue), neutral (green) and molecular (red)
    material.}
\label{fig:nB_Bstar}
\end{figure}

\begin{figure}
\includegraphics[width=\linewidth]{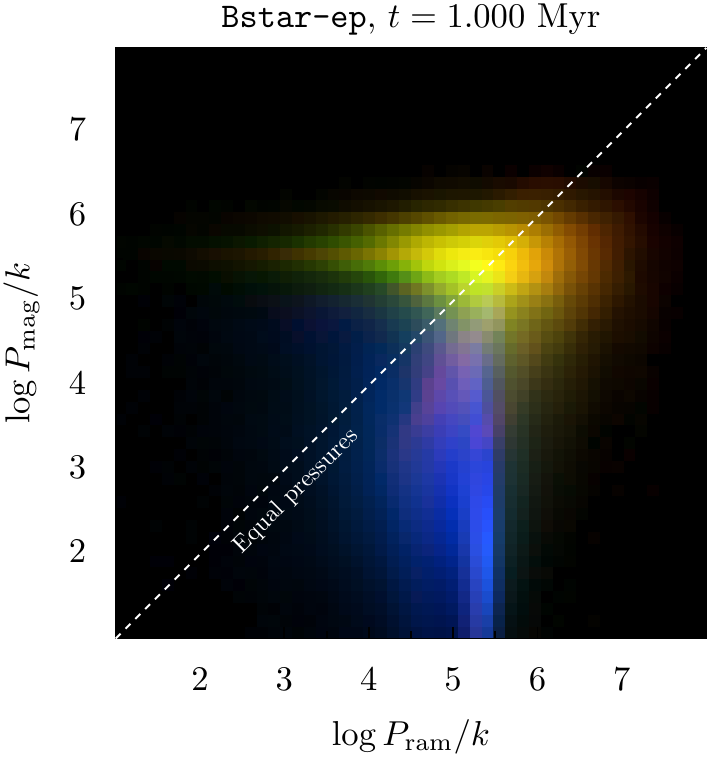}
  \caption{Magnetic pressure versus turbulent pressure for
    volume-weighted joint distributions in the case of the B-star at 
    $t=10^6$~years. Note that the scales are the same on both $x$ and
    $y$ axes.  The diagonal line shows equal pressures. The three
    colours represent ionised (blue), neutral (green) and molecular
    (red) material.}
\label{fig:pram_pmag_Bstar}
\end{figure}
Even though we find that the presence of a magnetic field only has
small effects on the growth of the \ion{H}{2} region, this does not
mean that the field is uninteresting or unimportant. For one thing,
the field is important in the still neutral medium, where it
constitutes a significant fraction of the total energy.

In Fig.~\ref{fig:cuts_Bstar} we show the magnetic field and the
distribution of the photoionised gas for the B star case in the same
format as Figs.~\ref{fig:krum1} and \ref{fig:krum2}. We can see that
the magnetic field looks very different. The initial conditions were
the result of an MHD turbulence calculation and in the molecular gas
(coloured grey/white in the lower panels of Fig.~\ref{fig:cuts_Bstar})
we can see that the field appears to be randomly oriented.

In the warm, neutral (PDR) gas (coloured purple in the lower panels of
the figure) there is strong evidence that the magnetic field is being
aligned parallel to the ionisation front, particularly in the lower
left region of the $x$-$y$ and $x$-$z$ plots. At some points in this
image, the ionisation front has left the grid, and the periodic
boundary condition in this simulation means that ionised material
exits one side of the grid and reappears at the opposite boundary,
where it will usually recombine because of the high column density
through the molecular and neutral gas to the star at this position. At
this point in the simulation, this is not affecting the global
evolution of the \ion{H}{2} region and surrounding warm, neutral gas.

The photoionised region has a very low magnetic field intensity, showing that
the field has been pushed aside by the expansion of the \ion{H}{2}
region. Photoevaporated flows from density inhomogeneities in both the
neutral and molecular gas drag the magnetic field with them into the
photoionised region in long filamentary structures. However, there is
little mass associated with these flows. In the $x$-$y$ and $x$-$z$
central planes there is a large region of partially ionised gas. This
is also in the direction where the ionisation front has left the grid
and could be due to a density or velocity gradient, which results in
the ionisation front being unable to keep up with the gas, since it is
no longer preceded by a neutral shock in this part of the
computational grid.

In order to analyze the role of the magnetic field in our three
components, ionised, neutral and molecular, we present colour plots
showing the joint distribution of various quantities in these three
components. Since the B-star and O-star results are quite comparable,
we only show the B-star results.

We start by considering joint distribution of the magnetic field
strength against the density.  Fig.~\ref{fig:nB_Bstar} shows this for
the case of the B-star at $t=10^6$~years. The distribution of the
ionised material is shown in blue and displays a wide range of field
strengths (0.1 to 10 $\mu$G) across a narrow range of densities (10 to
20 cm$^{-3}$; in the case of the O-star these values are some 5 times
higher). The diagonal lines show the Alfv\'en speeds. If the sound
speed in the gas is smaller than $v_\mathrm{A}$ the flow will be
magnetically dominated. As can be seen, the ionised material does not
reach Alfv\'en speeds of 10~km~s$^{-1}$, the typical sound speed in
the ionised medium. Consequently the ionised flow is never
magnetically dominated, consistent with the results found in the
previous sections.

Another way to represent the role of the magnetic field is to plot the
joint distribution of magnetic and ram (turbulent) pressure, as is
done in Fig.~\ref{fig:pram_pmag_Bstar}. In full equipartion, the
turbulent, magnetic and thermal pressures should all be equal. Since
the \ion{H}{2} region is expanding through its thermal pressure, it is
the thermal pressure which is the overall dominating
one. Fig.~\ref{fig:pram_pmag_Bstar} shows that of the other two
pressure components, turbulent pressure mostly dominates over magnetic
pressure. In only a few places in the \ion{H}{2}
region the magnetic pressure dominates. 

In the neutral and molecular material (represented by green and red in
the joint distribution figures) the situation is different. The joint
distribution of density and magnetic field strength,
Fig.~\ref{fig:nB_Bstar}, shows a wide range of densities ($10^2$--$10^4$~cm$^{-3}$) and a narrower range of magnetic field strengths
(mostly around 10--30~$\mu$G, although some areas, principally neutral
ones, have magnetic fields as weak as 1~$\mu$G). The molecular material
generally has higher densities than the neutral material. Since the sound
speed in the neutral and molecular regions is 1~km~s$^{-1}$ or less,
they are mostly magnetically rather than thermally dominated. In the
distribution of pressures, Fig.~\ref{fig:pram_pmag_Bstar}, one sees
that the neutral and molecular material clusters around equipartition
of the magnetic and turbulent pressures. The molecular material has
regions where the turbulent pressure dominates, and others where the
magnetic pressure dominates. The neutral (PDR) material has generally
somewhat lower pressures and is closer to equipartition.

Fig.~\ref{fig:nB_Bstar} can be compared to observationally derived
values for the magnetic field and densities in \ion{H}{2} regions and their
surroundings. Results of Harvey--Smith et al. (in preparation) on large
Galactic \ion{H}{2} regions show ranges in magnetic field strengths and typical
densities for the ionised, neutral and molecular components very comparable
to what we find in our simulation results. 

\section{Discussion}
\label{sec:discussion}
\subsection{Comparison with observations: RCW\,120}
\label{subsec:rcw120}
The \ion{H}{2} region RCW\,120 and its surrounding medium have
recently been extensively studied at near- to far-infrared wavelengths
in the context of triggered star formation
\citep{{2007A&A...472..835Z},{2009A&A...496..177D},{2010A&A...518L..99A},{2010A&A...510A..32M},{2010A&A...518L..81Z}}.
This small \citep[3.5~pc diameter at 1.35~kpc distance,
][]{2007A&A...472..835Z} \ion{H}{2} region appears to have a single
ionising source, identified from VLT-SINFONI near-IR spectroscopy and
spectral line fitting to stellar atmosphere models as an O6--O8V/III
star \citep{2010A&A...510A..32M} with an effective temperature of
$T_\mathrm{eff} = 37.5\pm2$~kK and ionising photon rate of $\log
Q_\mathrm{H} = 48.58\pm0.22$. These observationally derived
parameters are remarkably similar to those we have adopted for our O
star simulations and so an opportunity for a direct comparison of
RCW\,120 with our results presents itself.

Morphologically, the observations and simulation look very similar. In
particular, our Fig.~\ref{fig:images_O}, corresponding to the O star
simulation after 200,000~yrs of evolution, is the same size as the
observed bubble RCW\,120. The upper panels of the figure show the
optical emission (with extinction from dust), while the lower panels
show synthetic 6\,cm radio free-free emission from the ionised gas
(blue), generic PAH emission from the PDR (green) and cold molecular
gas column density (red). Figs.~1 and 2 of
\citet{2009A&A...496..177D} show the H$\alpha$ emission from the
ionised gas, the $8\,\mu$m emission from PAHs in the PDR, $24\,\mu$m
emission from warm dust in the photoionised region, and
$870\,\mu$m cold dust emission from the neutral and molecular gas.
If the reader mentally rotates the simulated images so that the lower
left corner moves to the top centre, then a certain correspondence
between simulations and observations can be imagined. The ionised gas
emission fills the central cavity. In the optical, the images appear
criss-crossed by dust lanes. Some of these are foreground, others are
the result of projection of filaments and ridges deeper inside the
\ion{H}{2} region. The PAH emission comes from a thin shell around the
periphery of the \ion{H}{2} region, with projection seeming to put
some PAH emission in the interior of the \ion{H}{2} region, as can be
seen in both the simulation and the observations. The simulated PAH
emission presents the same sort of irregularities, filaments and
clumps as are seen in the observations.

The initial total mass in the computational box is $\sim
2000\,M_{\odot}$. After 200,000~yrs of evolution, the mass
distribution, as seen in Fig.~\ref{fig:comp2_O}, is approximately
5\% ionised gas, 55\% neutral gas, and 40\% molecular gas. Some of the
molecular gas is undisturbed material at the edge of the computational
box but quite a large proportion must be in the clumps and filaments
in the \ion{H}{2} region and PDR, where it becomes compressed by
radiation-driven implosion. The \ion{H}{2} region has influenced
virtually all of the computational domain by this time, and the $\sim
2000\,M_{\odot}$ of neutral and molecular material is comparable to
the 1100--$2100\,M_{\odot}$ derived by \citet{2007A&A...472..835Z} and
\citet{2009A&A...496..177D}
from the $870\,\mu$m cold dust emission. This dust will be present in
both the neutral and molecular gas. Hence, we can surmise that the
average density in the vicinity of RCW\,120 is similar to that in our
computational box, that is $\langle n_0 \rangle = 10^3$~cm$^{-3}$,
though, of course, the density distribution is far from
homogeneous. From this, we could go so far as to postulate that the
age of RCW\,120 must be similar to the time depicted in our
simulation, that is, of order 200,000~yrs. \citet{2010A&A...510A..32M}
were unable to assign an age to RCW\,120 except to say that it must be
younger than 5~Myr.

Although radiation-driven implosion does enhance the density
of photoionised globules at the periphery of the simulated \ion{H}{2}
region, the fact that we do not include self gravity in our
simulations means that we are unable to model
triggered star formation in the neutral shell around
the \ion{H}{2} region. However, it is quite likely that
radiation-driven implosion could trigger gravitational collapse in
clumps that are already on the point of forming stars.

There are other potentially important physical processes that we do
not include in our models. Dust grains are an important component of
the interstellar medium in star-forming regions. They absorb ionising
photons and both observations (e.g., \citealp{1989ApJS...69..831W})
and theory (e.g., \citealp{2004ApJ...608..282A}) suggest that more
than 50\% of all the ionizing photons are absorbed by dust within the
\ion{H}{2} regions in the initial stages of evolution when the ionised
density is high. Radiation pressure on dust grains can form a central
cavity in the \ion{H}{2} region \citep{{1967ApJ...147..965M},
  {2002ApJ...570..688I}}, which can be as large as 30\% of the
ionisation front radius. Such cavities have been known for a long time
from optical observations of \ion{H}{2} regions (e.g.,
\citealp{1962ApJ...135..394M}) and are also seen in more recent
infrared observations (e.g.,
\citealp{2008ApJ...681.1341W}). \citet{2009ApJ...703.1352K} have
studied the effect of radiation pressure on the dynamics of \ion{H}{2}
regions and conclude that it is only an important effect for massive
star clusters and not for \ion{H}{2} regions around individual massive
stars. Stellar winds are also to be expected from stars whose
effective temperatures are greater than 25,000~K and these winds will
provide an alternative mechanism for evacuating a central cavity in
the dust and gas distribution in the \ion{H}{2} region. The mechanical
luminosity of the stellar wind is converted into thermal pressure by
shock waves \citep{1997pism.book.....D} and this could affect the
dynamics of the \ion{H}{2} region.

\subsection{Predicted maps of projected magnetic field}
\label{sec:pred-maps-proj}
\begin{figure*}
  \includegraphics{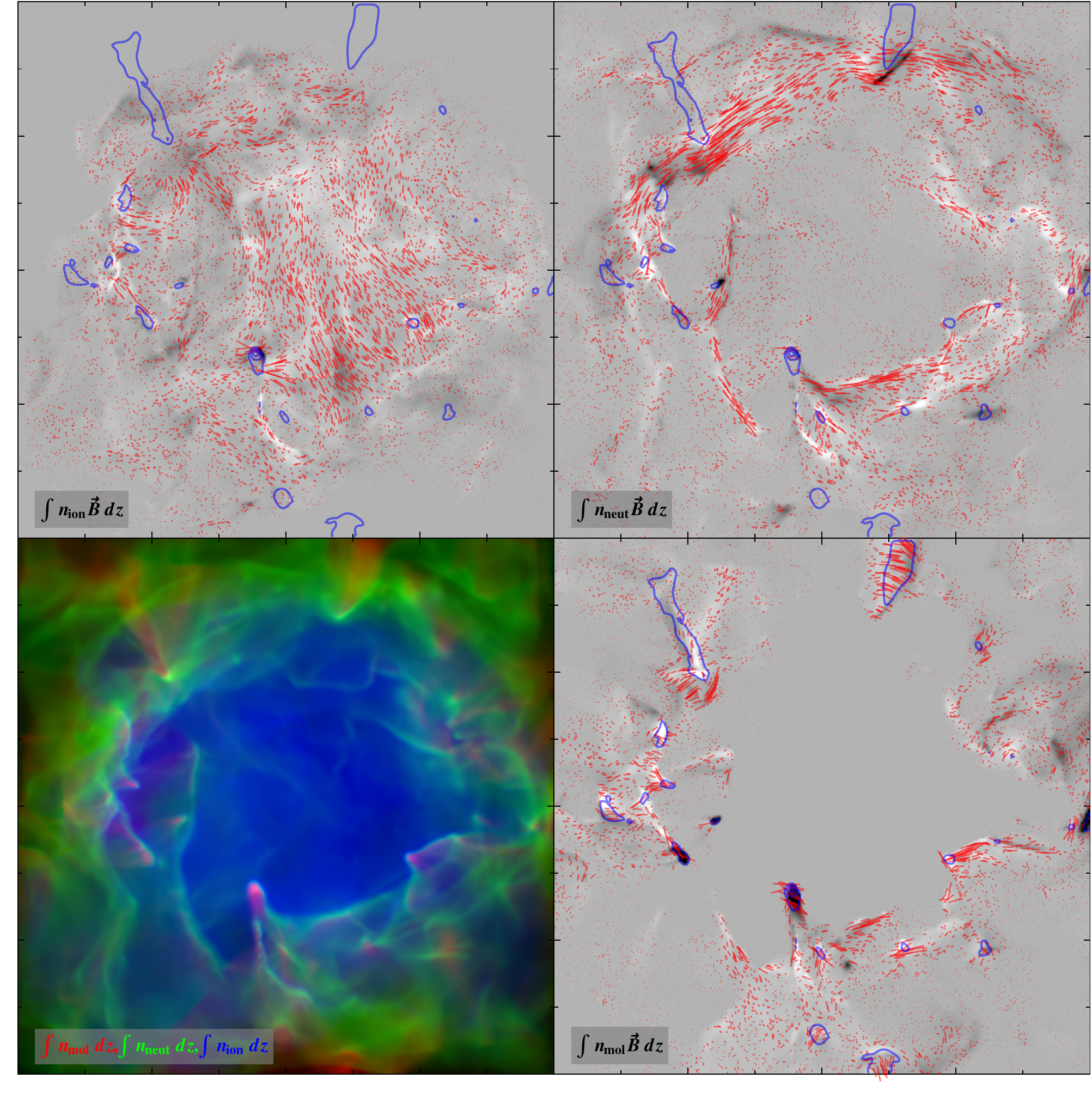}
  \caption{Projected magnetic field of O star model at a time of \(t =
  200,000\)~years, showing the full simulation cube (extent of 4~pc
  along each axis). The horizontal and vertical graph axes correspond
  to the simulation \(x\) and \(y\) axes, respectively. The three
  grayscale images (top-left, top-right, and bottom-right panels) show
  the line-of-sight integral of the line-of-sight component of the
  magnetic field, weighted by the ionised (top-left), neutral
  (top-right), or molecular (bottom-right) gas density. The
  superimposed red vectors indicate the magnitude and direction of the
  line-of-sight integral of the plane-of-sky components of the
  magnetic field, using the same weightings. The integrals of the
  plane-of-sky components were carried out in a Stokes \(QU\) frame,
  as described in the appendix. Blue contour lines show the column
  density of molecular gas, to aid comparison between the panels. The
  bottom-left panel shows a composite image of the column densities of
  molecular (red), neutral(green), and ionised (blue) gas. The
  normalisation varies between the panels. For the column densities
  (bottom-left panel and contours) the maximum plotted values in units
  of H nuclei per cm\(^2\) are \(3.7\times 10^{21}\) (ionised),
  \(5.1\times 10^{22}\) (neutral), and \(1.3 \times 10^{23}\)
  (molecular). For the projected magnetic fields, the maximum plotted
  values in units of \(\mu\)G times H nuclei per cm\(^2\) are \(5.2
  \times 10^{22}\) (ionised, top left), \(2.1\times 10^{24}\)
  (neutral, top right), \(1.7\times 10^{24}\) (molecular, bottom
  right).}
  \label{fig:bproj-ostar-full}
\end{figure*}
\begin{figure*}
  \includegraphics{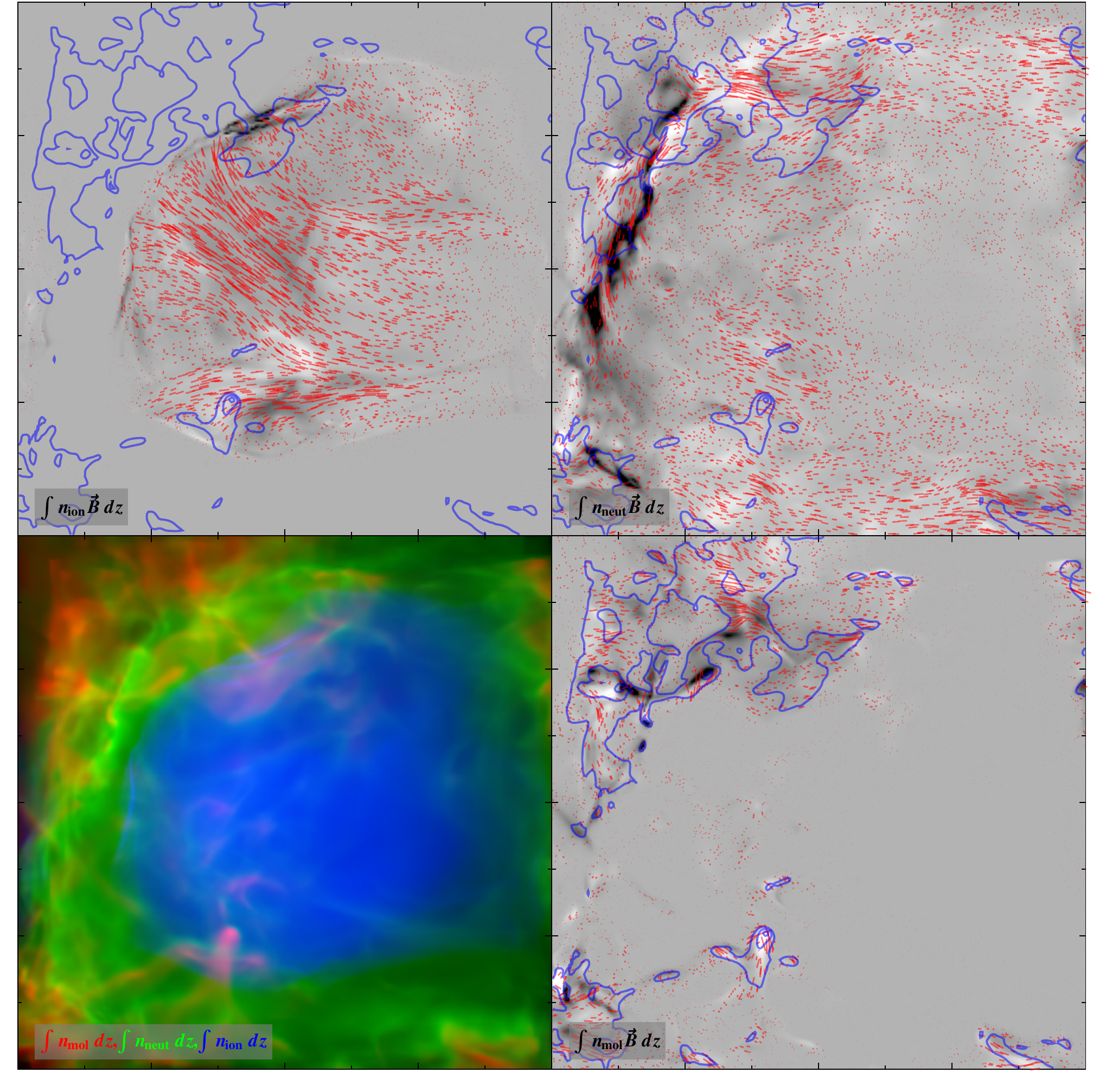}
  \caption{Projected magnetic field of B star model at a time of \(t =
  10^6\)~years, showing the full simulation cube (extent of 4~pc
  along each axis). Panels as in Fig.~\ref{fig:bproj-ostar-full}. For
  the column densities (bottom-left panel and contours) the maximum
  plotted values in units of H nuclei per cm\(^2\) are \(2.1\times
  10^{20}\) (ionised), \(4.6\times 10^{22}\) (neutral), and \(3.4
  \times 10^{22}\) (molecular). For the projected magnetic fields, the
  maximum plotted values in units of \(\mu\)G times H nuclei per
  cm\(^2\) are \(1.0 \times 10^{21}\) (ionised, top left), \(7.5\times
  10^{23}\) (neutral, top right), \(7.2\times10^{23}\) (molecular,
  bottom right).}
  \label{fig:bproj-bstar-full}
\end{figure*}
\begin{figure*}
  \includegraphics{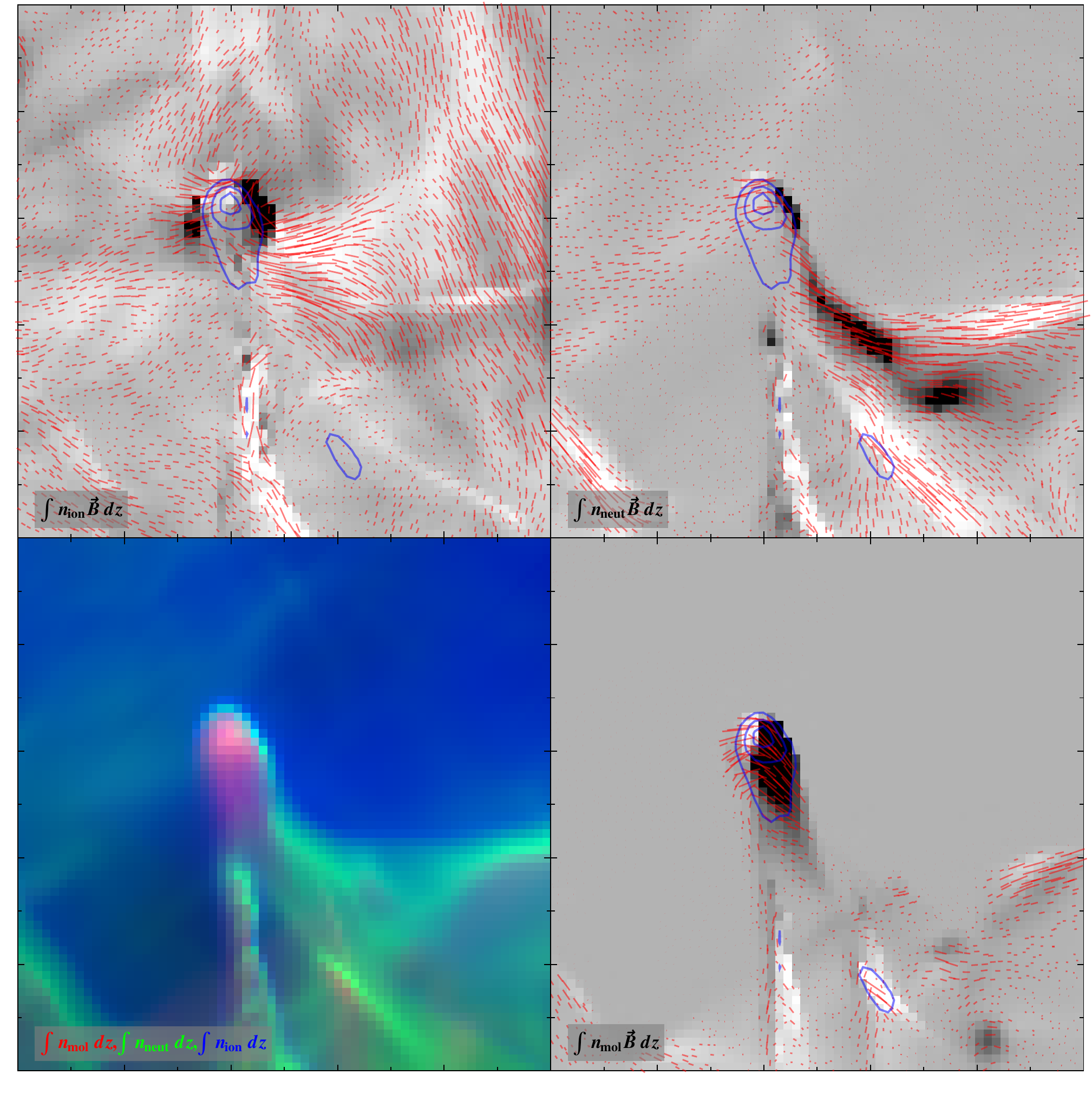}
  \caption{Projected magnetic field of O star model at a time of \(t =
  200,000\)~years, showing a \(4\times\) zoom centred on the southern
  dense globule (extent of 1~pc along each axis). Panels as in
  Fig.~\ref{fig:bproj-ostar-full}. For the column densities
  (bottom-left panel and contours) the maximum plotted values in units
  of H nuclei per cm\(^2\) are \(3.7\times 10^{21}\) (ionised),
  \(2.7\times 10^{22}\) (neutral), and \(1.3 \times 10^{23}\)
  (molecular). For the projected magnetic fields, the maximum plotted
  values in units of \(\mu\)G times H nuclei per cm\(^2\) are \(2.6
  \times 10^{22}\) (ionised, top left), \(1.0\times 10^{24}\)
  (neutral, top right), \(2.2\times 10^{24}\) (molecular, bottom
  right).}
  \label{fig:bproj-ostar-globule}
\end{figure*}
\begin{figure*}
  \includegraphics{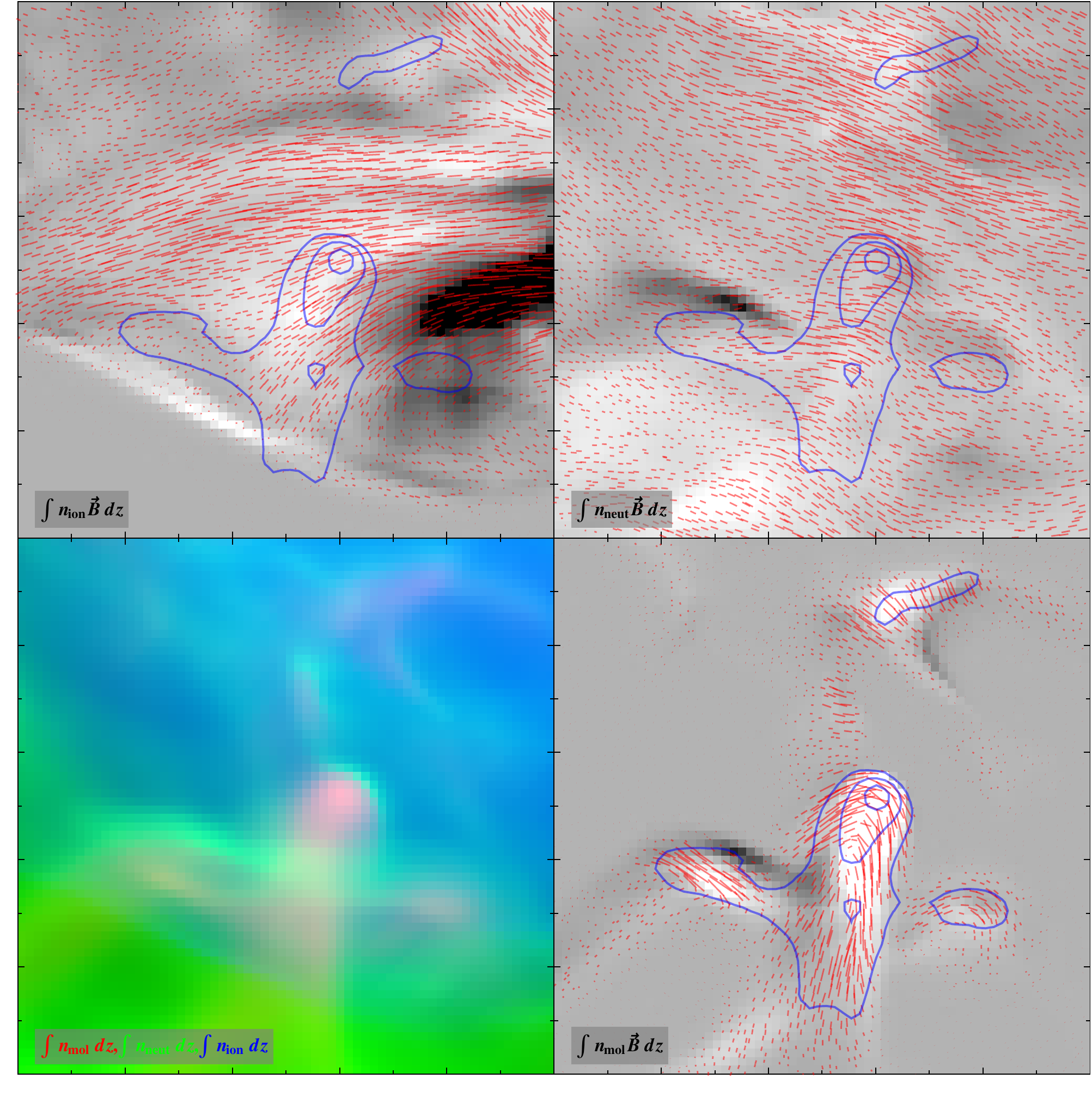}
  \caption{Projected magnetic field of B star model at a time of \(t =
  10^6\)~years, showing a \(4\times\) zoom centred on the
  southern dense globule (extent of 1~pc along each axis). Panels as
  in Fig.~\ref{fig:bproj-ostar-full}. For the column densities
  (bottom-left panel and contours) the maximum plotted values in units
  of H nuclei per cm\(^2\) are \(1.7\times 10^{20}\) (ionised),
  \(1.5\times 10^{22}\) (neutral), and \(3.2 \times 10^{22}\)
  (molecular). For the projected magnetic fields, the maximum plotted
  values in units of \(\mu\)G times H nuclei per cm\(^2\) are \(5.1
  \times 10^{20}\) (ionised, top left), \(3.8\times 10^{23}\)
  (neutral, top right), \(3.6\times 10^{23}\) (molecular, bottom
  right).}
  \label{fig:bproj-bstar-globule}
\end{figure*}
Figs.~\ref{fig:bproj-ostar-full} to \ref{fig:bproj-bstar-globule}
show visualisations of the projected integrated B-field (line-of-sight
and plane-of-sky components) from our simulations. These
visualisations represent generic idealised versions of maps that can
be obtained from various observational techniques. The line-of-sight
field component can be determined from the Faraday rotation measure
\citep{1981ApJ...247L..77H} for ionised gas or from Zeeman
spectroscopy \citep{2001ApJ...560..821B} for neutral/molecular
gas. The plane-of-sky field components can be determined from
observations of polarized emission or absorption
\citep{2008AJ....136..621K} from aligned spinning dust grains. In all
of these techniques, the determination of \(\mathbf{B}\) is not
straightforward and relies on many auxiliary assumptions. In
particular, it is very difficult to determine the absolute magnitude
of the plane-of-sky field components except via statistical techniques
such as the Chandrasekhar-Fermi method
\citep{{1953ApJ...118..113C},{2001ApJ...561..800H},{2001ApJ...546..980O},{2008ApJ...679..537F}}. Furthermore, many
of the techniques do not uniformly sample all regions along the line
of sight, but are biased towards regions with particular physical
conditions.  We therefore caution against direct comparison of
observational results with our maps in any but a qualitative
sense. The maps are nonetheless useful in showing an overview of the
magnetic field geometry in our simulations. In the following
discussion, we describe locations on the projected map using the
points of the compass, assuming that north is up (positive \(y\)) and
east is left (negative \(x\)).

The most striking aspect of Figs.~\ref{fig:bproj-ostar-full} and
\ref{fig:bproj-bstar-full} is the large-scale order that is apparent
in the projected magnetic field. This is particularly visible in the
neutral-weighted maps (top-right panels), where it can be seen that
the field is frequently oriented parallel to the large-scale
ionisation front, forming a ring around the \hii{} region. The effect
is particularly strong along the north and south borders of the region
because of the net positive field along the \(x\)-axis (see
\S~\ref{sec:setup}), but it is also seen to a lesser extent along the east and
west borders, despite the fact that the mean \(y\)-component of the
field is zero. This is because of the fast-mode MHD shock that is
driven into the surrounding gas by the expanding \hii{} region and
which compresses both the gas and the field, tending to bend the field
lines so that they are more closely parallel to the shock than they
were in the undisturbed medium. A large-scale pattern is much harder
to discern in the molecular-weighted maps (bottom-right panels),
partly because the molecular column density is much less smoothly
distributed, being concentrated in globules and filaments.

In most dense filaments, the field direction is parallel to the long
axis of the filament, as can be seen particularly clearly in
Fig.~\ref{fig:bproj-ostar-globule}, which shows a detail view of the
dense photoevaporating globule found to the south in
Fig.~\ref{fig:bproj-ostar-full} and which is fed by multiple
neutral/molecular filaments. In the molecular gas at the head of the
globule, the magnetic field is bent into a hairpin shape. The B-field
in the ionised gas at the head of the globule tends to be oriented
perpendicular to the ionisation front, as is the case with almost all
the dense globules visible in Fig.~\ref{fig:bproj-ostar-full}. The
same is true along much of the bar-like feature to the west of the
globule. On the other hand, in a few regions, such as along the
filament that extends south from the globule, the B-field in the
ionised gas lies along the ionisation front.

Fig.~\ref{fig:bproj-bstar-globule} shows a detail view of the same
dense southern globule, but from the B star simulation shown in
Fig.~\ref{fig:bproj-bstar-full}.  The molecular gas shows a similar
field pattern to that seen in the O star case: the field goes up one of the
feeding filaments and down the other, with a hairpin bend in
between. Although this can be seen clearly in the molecular gas, the
neutral and ionised gas show very different patterns, with magnetic
field vectors that are generally perpendicular to the filament and
continuous with the large-scale field pattern in the region. This is
because the total ionised and neutral columns are dominated by diffuse
material along the line of sight, rather than by material associated
with the globule and filament.

\subsection{Effects of magnetic field on globule formation and evolution}
\label{sec:effects-magn-field}
 
It is interesting to compare the properties of the globules generated
in our turbulent simulations with the results of previous detailed
studies of the photoionisation of isolated dense globules
\citep{{2007Ap&SS.307..179W}, {2009MNRAS.398..157H}, {2010arXiv1012.1500M}}. The principal
findings of the earlier studies are that a sufficiently strong
magnetic field (\(\beta < 0.01\) in the initial neutral globule)
will produce important qualitative changes in the photoevaporation
process. Depending on the initial field orientation with respect to
the direction of the ionising photons, either extreme flattening of
the globule may occur (perpendicular orientations) or the radiative
implosion of the globule may be prevented (parallel orientations). For
all orientations the ionised photoevaporation flow cannot freely
escape from the globule, leading to recombination at late times. On
the other hand, a weaker field (\(\beta \simeq 0.01\) in the initial
globule) leads only to a moderate flattening of the globule and does
not prevent the free escape of the ionised photoevaporation flow.
 
In our turbulent simulations, the mean initial value of \(\beta\) is
\(\simeq 0.032\) (\S~\ref{sec:setup}), which is intermediate between
the two cases discussed above and might lead one to suspect that
magnetic effects on the evolution of globules should be
substantial. However, this is \emph{not} the case. A careful
examination of the three-dimensional globule morphologies in our
simulations shows no evidence of magnetically induced
flattening. Although many globules do show asymmetries, this is true
equally of our non-magnetic simulations and is presumably due to their
irregular initial shape and internal turbulent motions. The only
difference in the globule properties between our non-magnetic and
magnetic simulations is that the very smallest globules do not seem to
form in the magnetic case.
 
In order to explain this apparent discrepancy with earlier work, it is
necessary to examine in more detail the distribution of magnetic field
in the initial conditions of our turbulent simulations. It turns out
that the dense filaments of molecular gas (from which the globules
will ultimately form) tend to be much less magnetically dominated than
the more diffuse gas, typically showing \(\beta >
0.1\). Furthermore, these filaments already show a magnetic geometry
similar to that described above (\S~\ref{sec:pred-maps-proj}), with the
field running along the long axis of the filament and with the field
changing sign as one moves across the short axis. This is very
different from the initial conditions assumed in the earlier globule
photoevaporation studies, which were a uniform magnetic field that
threaded a spherical globule (e.g., Fig.~1 of
\citealp{2009MNRAS.398..157H}). As a result, the compressed globule heads in
our turbulent simulations show approximately equal thermal and
magnetic pressures (\(\beta \sim 1\)) and the magnetic effects are
very modest. One caveat to this result is that numerical diffusion due
to our limited spatial resolution may be producing non-physical
magnetic reconnection in the globule head when oppositely directed
field lines are forced together. Higher resolution studies of globule
implosion with realistic initial field configurations are required in
order to clarify this \NEW{and these should also include self gravity, which may be important during the phase of maximum compression \citep{2007MNRAS.377..383E}}.

\section{Conclusions}
\label{sec:conclusion}
We have performed radiation-magnetohydrodynamic simulations of the
formation and expansion of \ion{H}{2} regions and PDRs around an O
star and a B star in a turbulent magnetised molecular cloud on scales
of up to 4~parsec. Our principal conclusions are as follows:
\begin{enumerate}
\item The expansion of the \ion{H}{2} region is little affected by the
presence of the magnetic field, since the thermal pressure of
the ionised gas dominates the dynamics on the timescales of our
simulations (\S~\ref{sec:glob-prop-prot}).
\item The O star simulations produce greater fragmentation and denser
clumps and filaments around the periphery of the \ion{H}{2} region
than the B star case (\S~\ref{sec:morph}, Figs.~\ref{fig:images_O} and
\ref{fig:images_B}).
\item For B stars the non-ionising far ultraviolet radiation plays an
important role in determining the morphology of the region. \ion{H}{2}
regions around such stars are surrounded by a thick, relatively smooth
shells of neutral material (PDR), $\sim 30\%$ of the bubble radius
(e.g., lower panels of Fig.~\ref{fig:cuts_Bstar}). In
the O star simulations, the PDR is thinner and more irregular in
shape (e.g., Fig.~\ref{fig:images_O}). 
\item The resemblance at optical and longer wavelengths of our
simulations to observed bubbles is striking
(\S~\ref{subsec:rcw120} and \citealp{2009A&A...496..177D,{2010A&A...523A...6D}}). Our $\sim 2$~pc radius bubbles are a
typical size compared to bubbles at known distances in the
GLIMPSE surveys \citep{2006ApJ...649..759C, 2007ApJ...670..428C},
and so comparisons are meaningful.

\item The expanding \ion{H}{2} region and PDR tend to erase
  pre-existing small-scale disordered structure in the magnetic field,
  producing a large-scale ordered field in the neutral shell, with
  orientation approximately parallel to the ionisation front (top
  panels of Fig.~\ref{fig:cuts_Bstar} and top-right panels of
  Figs.~\ref{fig:bproj-ostar-full} and \ref{fig:bproj-bstar-full}).

\item Dense evaporating globules, pillars, and elephant trunk
  structures tend to be fed by two or more neutral/molecular
  filaments, with magnetic fields running along their length (right
  panels of Fig.~\ref{fig:bproj-ostar-globule} and bottom-right panel
  of Fig.~\ref{fig:bproj-bstar-globule}). The field geometry in the
  neutral and molecular gas at the bright head of the structure tends
  to be of a hairpin shape.

\item The weak magnetic field in the ionised gas also shows an ordered
  structure. On the largest scales and at the latest times, it tends
  to align (albeit weakly in the O star case), with the mean field
  direction of the simulation's initial conditions
  (Fig.~\ref{fig:bproj-bstar-full}, top-left panel). On smaller scales, there is a
  general (but not universal) tendency for the field in ionised gas to
  be oriented perpendicular to the local ionisation front. This
  tendency is more pronounced in the O star simulation
  (Fig.~\ref{fig:bproj-ostar-full}, top-left panel) and in globule evaporation flows
  for both simulations (top-left panels of Figs.~\ref{fig:bproj-ostar-globule} and
  \ref{fig:bproj-bstar-globule}).

\item The highest pressure compressed neutral/molecular gas shows
  approximate equipartition between thermal, turbulent, and magnetic
  energy density, whereas lower pressure gas (either neutral or
  molecular) tends to separate into, on the one hand, magnetically
  dominated, quiescent regions, and, on the other hand, demagnetised,
  highly turbulent regions (\S~\ref{sec:magnetic-quantities},
  Fig.~\ref{fig:pram_pmag_Bstar}). 
  The lower pressure gas also
  separates into low-\(\beta\), magnetically dominated regions (which
  are largely molecular) and high-\(\beta\), thermally dominated
  regions (which are largely neutral). The ionised gas, on the other
  hand, always shows approximate equipartition between thermal and
  turbulent energies, but with the magnetic energy being lower by 1 to
  3 orders of magnitude.

\item Velocity dispersions in the ionised gas of \(7\)--\(9\)~\kms{}
  are maintained for the entire duration of all our simulations
  (\S~\ref{sec:glob-prop-prot}, Fig.~\ref{fig:comp3_O} and
  \ref{fig:comp3_B}). This is 5 to 10 times higher than the value that
  would be predicted by expansion in a uniform medium. At early times
  (\(t < 100,000\)~yr for the O star, or \(t < 300,000\)~yr for the B
  star), this dispersion is mainly due to radial champagne-flow
  expansion as the \hii{} region escapes from its natal clump. At
  later times, the net radial expansion of ionised gas subsides, but
  the velocity dispersion is maintained by inwardly-directed
  photoevaporation flows from globules and pillars.


\end{enumerate}

\section*{Acknowledgments}
SJA and WJH acknowledge financial support from DGAPA PAPIIT projects
IN112006, IN100309 and IN110108. This work was supported in part by
Swedish Research Council grant 2009-4088. Some of the numerical
calculations in this paper were performed on the Kan Balam
supercomputer maintained and operated by DGSCA, UNAM. This work has
made extensive use of NASA's Astrophysics Abstract Data Service and
the astro-ph archive.

\appendix 
\section{Integration of plane-of-sky field components}
Polarisation-based methods for measuring the plane-of-sky components
of the magnetic field cannot distinguish the sense of the field and
are degenerate between a position angle \(\theta\) and a position
angle \(\theta + 180^\circ\). It is therefore not sufficient to simply
integrate \(B_x\) and \(B_y\) along the line of sight with appropriate
weighting. Neither is it sufficient to integrate \(B_x^2\) and
\(B_y^2\), since this will always give a result in the first quadrant
(\(B_x, B_y > 0\)). In order to calculate our projected magnetic field
maps (\S~\ref{sec:pred-maps-proj}), we therefore adopt a ``Stokes
parameter'' approach \citep{Chandrasekhar:1960}, whereby the
plane-of-sky components \(B_x(x,y,z)\) and \(B_y(x,y,z)\) are first
transformed to (\(Q, U\))-space as \(B_Q = \tilde{B} \cos 2 \theta\),
\(B_U = \tilde{B} \sin 2 \theta\), where \(\tilde{B} = (B_x^2 +
B_y^2)^{1/2}\) and \(\theta = \tan^{-1} (B_y/B_x)\). Then the
line-of-sight integration is performed as \(M_{Q,U}(x,y) = \int
B_{Q,U}(x,y,z)\, n \, dz\), where \(n\) is the relevant density
(ionised, neutral, or molecular). Finally, the components of the
projected field are transformed back to (\(x,y\))-space: \(M_x =
\tilde{M} \cos\theta\), \(M_y = \tilde{M} \sin\theta\), where
\(\tilde{M} = (M_Q^2 + M_U^2)^{1/2}\) and \(\theta = 0.5 \tan^{-1}
(M_U/M_Q)\).
\end{document}